%% file: main.tex
\documentclass{article}

\usepackage{PRIMEarxiv}

\usepackage[utf8]{inputenc} 
\usepackage[T1]{fontenc}    
\usepackage{hyperref}       
\usepackage{url}            
\usepackage{natbib}
\usepackage{booktabs}       
\usepackage{amsfonts}       
\usepackage{nicefrac}       
\usepackage{microtype}      
\usepackage{xcolor}         
\usepackage{enumitem}
\usepackage{amssymb}
\usepackage{pifont}
\usepackage{diagbox}
\usepackage{multiaudience}

\usepackage{amssymb}
\usepackage[small]{caption}
\usepackage{graphicx}
\usepackage{amsmath}
\usepackage{amsthm}
\usepackage{booktabs}
\usepackage{algorithm}
\usepackage{algorithmic}
\usepackage{amsmath} 
\usepackage{subfigure}
\usepackage{multirow}
\usepackage{booktabs}
\usepackage{wrapfig}

\title{MoFa: A Unified Performance Modeling Framework for LLM Pretraining}

\author{%
  AIH Training Team
}

\begin{document}

\noindent  
\begin{minipage}{0.08\textwidth} 
    \centering
    \includegraphics[width=1.5cm]{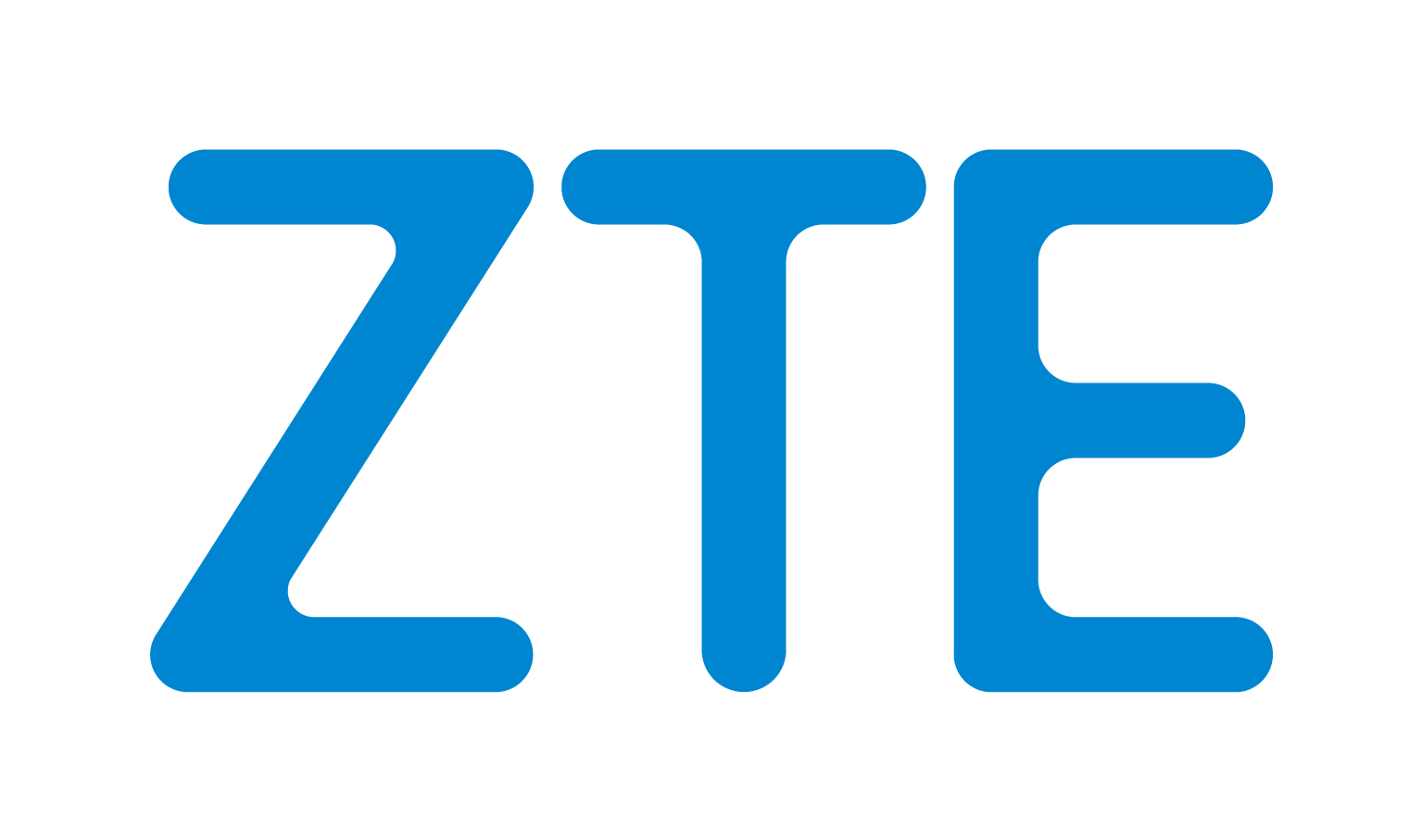}  
\end{minipage}
\hfill 
\begin{minipage}{0.08\textwidth} 
    \centering
    \includegraphics[width=1.05cm]{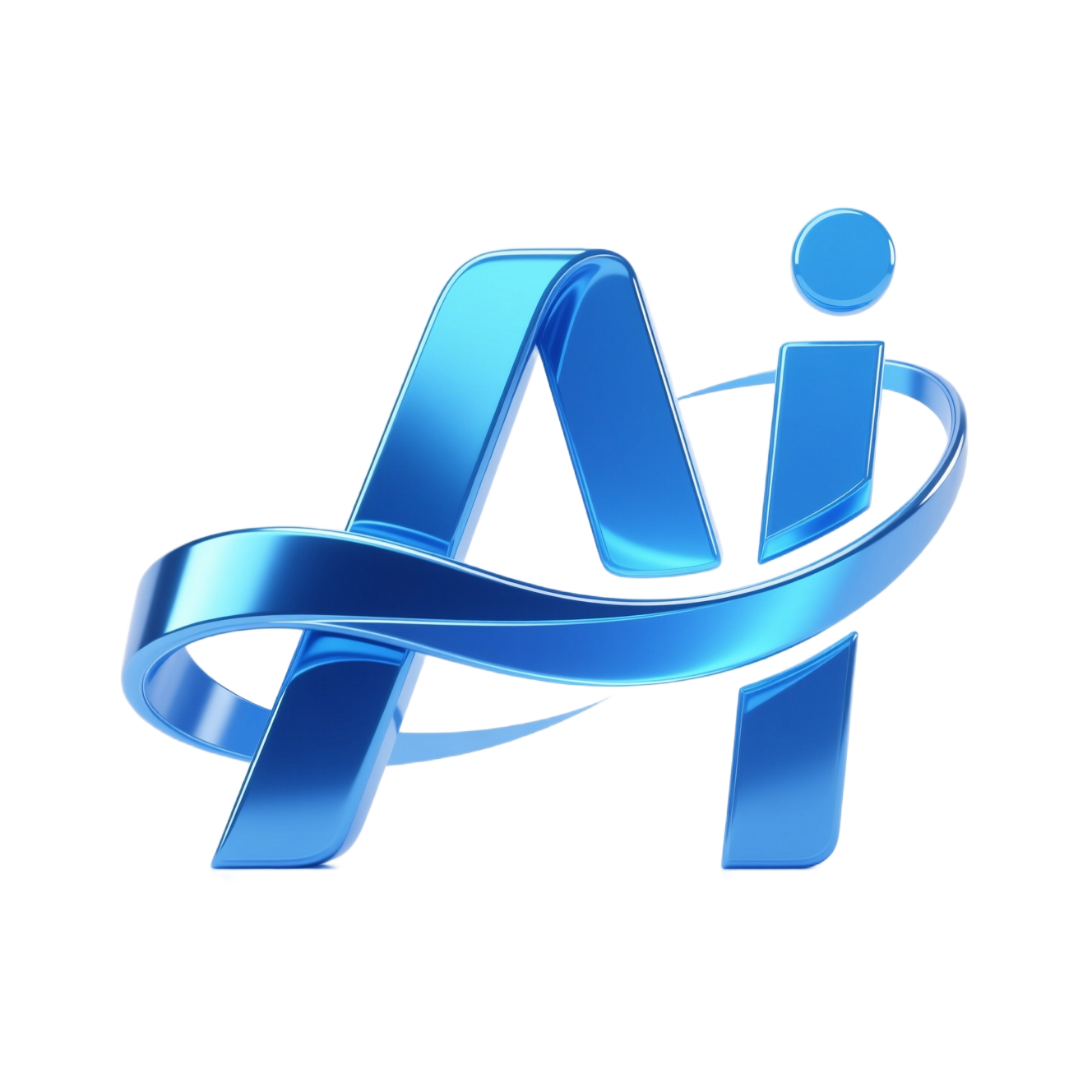} 
\end{minipage}
\vspace*{-0.3cm}
\maketitle

\begin{abstract}
The exponential growth in LLM scales, with parameters soaring from billions to trillions, has necessitated distributed pretraining across large clusters comprising thousands to tens of thousands of devices. While hybrid parallelization strategies enable such pretraining, the vast combinatorial strategy space introduces significant optimization challenges. Traditional manual tuning methods incur prohibitive trial-and-error costs, and existing performance modeling approaches exhibit critical limitations: they fail to comprehensively account for prevalent optimization features and ignore the substantial overhead imposed by essential fault tolerance mechanisms like checkpoint recovery in long-duration pretraining. To address these gaps, we propose \textbf{MoFa}, a novel pretraining performance modeling framework that unifies \textbf{m}ulti-dimensional \textbf{o}ptimization features and \textbf{fa}ult tolerance. MoFa incorporates an enhanced cost model to accurately capture the effects of key optimizations and integrates a fault tolerance model based on historical cluster reliability data. Besides, a MoFa-based tuning system is developed to explore optimal pretraining performance and potential bottlenecks in various scenarios. Extensive modeling evaluations demonstrate that MoFa can achieve high prediction accuracy across various scenarios. In addition, through comprehensive tuning experiments, our framework systematically reveals the key factors influencing pretraining performance under different configurations, which provides solid a priori guidance for LLM pretraining system design and deployment.
\end{abstract}

\input{Introduction.tex}
\input{relatedwork.tex}
\input{mofadesign.tex}
\input{autotuningsys.tex}
\input{experiment.tex}
\input{futurework.tex}
\input{conclusion.tex}
\input{author.tex}

{\small
\bibliographystyle{unsrt}%
\bibliography{nips21}
}

\end{document}

%% file: Introduction.tex
\section{Introduction}
In recent years, with the widespread application of large language models (LLMs) such as GPT~\cite{achiam2023gpt}, LLaMA~\cite{grattafiori2024llama}, and DeepSeek~\cite{liu2024deepseek,guo2025deepseek} in natural language processing, multimodal reasoning~\cite{su2024federated,guo2022visually,li2024images}, code generation~\cite{shao2024deepseekmath}, and other domains~\cite{yin2022deep,zhang2023dual,zhang2022data}, there has been a dramatic increase in demand for large-scale, high-capability foundation models across various industries~\cite{su2024fedra}. Model scales continue to expand, with parameters rapidly escalating from tens of billions to trillions, while pretraining data volumes and computational requirements grow exponentially. Scaling Laws further indicate that increasing model parameters significantly enhances model capabilities~\cite{kaplan2020scaling}, continuously driving model scales toward higher magnitudes. However, the pretraining of ultra-large-scale models has far exceeded the computational and memory capacities of individual machines, and traditional small-to-medium-scale clusters are inadequate to meet these demands. Consequently, reliance on distributed clusters comprising thousands, tens of thousands, or even more Graphics Processing Units (GPUs) for collaborative pretraining is imperative~\cite{lin2017deep,shoeybi2019megatron}.

In large-scale clusters, hybrid parallel strategies have become core enabling technologies for large model pretraining~\cite{shoeybi2019megatron}. Nevertheless, the combinatorial space of parallel strategies in large-scale clusters is extremely vast~\cite{zheng2022alpa}. The selection of pretraining strategy not only impacts pretraining efficiency but also involves complex multi-dimensional constraints such as memory allocation and communication overhead~\cite{cui2018fast}. Traditional manual tuning methods relying on expert experience incur high trial-and-error costs and struggle to achieve global optima~\cite{liu2025galvatron}. Moreover, the long development cycles and substantial resource investments required for ultra-large-scale clusters make accurate performance prediction and strategy evaluation prior to actual deployment particularly critical~\cite{northrop2006ultra}. Therefore, constructing performance models capable of precisely modeling cluster behavior and supporting automated strategy optimization has emerged as a central challenge in large-scale pretraining systems~\cite{zheng2022alpa}.

Current mainstream modeling approaches in the industry primarily fall into two categories. On the one hand, white-box simulation methods are widely applied, which decompose and integrate computational, communication, and memory modules, combined with dynamic programming to search for optimal strategies~\cite{zheng2022alpa, li2024automatically, jia2019beyond, yuan2024accelerating}. On the other hand, the trace-driven method is proposed, which relies on runtime traces to construct fine-grained execution graphs for replaying actual execution processes~\cite{liu2025galvatron,liang2025lumos}. Although these methods exhibit respective advantages in strategy search efficiency or performance estimation, they lack consideration in the following two aspects:

\textbf{Various Optimization Features}: Current methods fail to comprehensively account for the impact of various optimization features widely adopted in practical distributed pretraining, like memory optimization and communication-computation overlap, leading to significant discrepancies between estimated and actual throughput.

\textbf{Fault Tolerance}: Current methods can not model dynamic factors in pretraining clusters, such as hardware, and do not incorporate the additional overhead introduced by fault tolerance mechanisms, like checkpoint recovery, making it difficult to accurately assess the effective performance of long-duration pretraining tasks.

To address these challenges, we propose \textbf{MoFa}, a performance modeling framework that unifies multi-dimensional optimization features and fault tolerance. MoFa incorporates an enhanced cost model to capture the impact of key optimizations on pretraining performance. Furthermore, it integrates a fault tolerance model based on historical cluster reliability data, enabling accurate characterization of pretraining behavior in large-scale GPU clusters with tens of thousands of devices. Building on this modeling foundation, we develop a tuning system that supports joint optimization of parallel strategies, optimization features, and fault tolerance methods for LLM pretraining tasks at various scales. This system provides predictive performance evidence and theoretical support for practical deployment. The main contributions of this work are summarized as follows:

\begin{itemize}
\item \textbf{A Multi-dimensional Optimization-Aware Performance Modeling Framework.}
The proposed framework incorporates diverse optimization features to significantly improve the accuracy of efficiency estimation in LLM pretraining scenarios.

\item \textbf{Historical Reliability-Aware Fault Tolerance Modeling.}
We are the first to integrate historical cluster reliability data into a performance modeling framework, enabling overhead estimation for typical fault tolerance scenarios and supporting end-to-end performance evaluation for long-duration pretraining tasks.

\item \textbf{A Unified Tuning System for Holistic Pretraining Configuration.}
We construct a performance tuning system that supports joint optimization of parallelization, optimization features, and fault tolerance strategies, offering a reliable prior-analysis foundation for system deployment.

\item \textbf{In-Depth Performance Analysis for Various-Scale Clusters.}
Through extensive modeling and tuning experiments, we systematically identify key performance bottlenecks in various-scale clusters including thousands even tens of thousands of GPUs, providing critical insights and theoretical guidance for future architecture design, optimization pathways, and scheduling strategies.
\end{itemize}

%% file: relatedwork.tex

\section{Related Work}
\subsection{Distributed Parallel Training Techniques}
In recent years, the successive releases of large-scale models have demonstrated that the scaling laws for large model training remain valid. As both model sizes and datasets continue to grow, the demand for computational power increases significantly. To support efficient model pretraining, the scale of large model training clusters has expanded from thousands to tens of thousands and even beyond tens of thousands of GPUs. Since the memory and computational capacity of a single GPU are insufficient to meet the requirements of pretraining models with hundreds of billions or even trillions of parameters, distributed parallel training techniques are essential to distribute the model and data across a large number of devices. Distributed parallel training serves as a core approach to address the memory wall and the computation wall of GPUs. Common distributed parallel training strategies are as follows:

\begin{itemize}
\item Data Parallelism (DP): The dataset is partitioned, and the model is replicated. Each DP group of devices holds a complete copy of the model and processes different data shards~\cite{rajbhandari2020zero}.
\item Tensor Parallelism (TP): The model is partitioned intra-layer, splitting large weight matrices across different devices for computation~\cite{shoeybi2019megatron}.
\item Sequence Parallelism (SP): The input sequence itself is partitioned and distributed across multiple computing devices for parallel processing~\cite{shoeybi2019megatron}.
\item Pipeline Parallelism (PP): The model is partitioned inter-layer, with different layers assigned to different devices. The computation process resembles a pipeline, and commonly used scheduling algorithms include GPipe~\cite{huang2019gpipe}, 1F1B~\cite{narayanan2021memory}, Interleaved 1F1B~\cite{narayanan2021efficient}, ZeroBubble~\cite{qi2023zero}, and DualPipe~\cite{liu2024deepseek}.
\item Expert Parallelism (EP): In Mixture-of-Experts (MoE) models, different experts are distributed across different devices~\cite{du2022glam}.
\item Context Parallelism (CP): Long input sequences are partitioned across different devices for computation, with results aggregated ultimately~\cite{liu2023ring,fang2024usp}.
\end{itemize}

In practical large-scale model training, multiple parallel strategies are almost always combined to achieve an optimal balance among memory usage, computation, and communication overhead. Particularly in ultra-large-scale clusters, hybrid distributed training has become an indispensable technique.

\subsection{Fault Tolerance Techniques}
Large-scale model pretraining is extremely time-consuming, often spanning weeks or months, and is typically conducted on clusters comprising tens of thousands of GPUs. Such extended operations are susceptible to interruptions caused by hardware failures~\cite{team2023gemini, lu2023multi} (e.g., GPU memory controller errors, NVMe SSD power loss), network anomalies~\cite{gupta2024evolution} (e.g., InfiniBand/RDMA interruptions), software exceptions~\cite{he2023unicron} (e.g., Out-of-Memory errors, NaN losses), and necessary human interventions (e.g., hyperparameter tuning, resource reallocation). For instance, Meta's research on pretraining the 405B parameter Llama 3 model highlights these challenges: the system operated on a cluster of 16,384 NVIDIA H100 GPUs and experienced an average of one failure every three hours, totaling 419 unexpected failures over 54 days. More than half of these incidents were attributed to GPU-related issues, particularly High Bandwidth Memory (HBM3). Due to the massive scale and highly synchronized nature of GPU training, the process is highly vulnerable to disruptions~\cite{wan2025robust}, where a single GPU failure can halt the entire training job, necessitating a restart.

Checkpointing and resuming training (CKPT)~\cite{lian2024universal} serves as the fundamental mechanism to ensure seamless recovery from interruptions in long-duration, large-scale AI training tasks. By systematically saving and restoring the complete training state (including model parameters, optimizer states, and random number generator seeds), it effectively mitigates the impact of hardware failures, network fluctuations, and manual interventions, thereby preventing significant wastage of computational resources and training progress.

In large-scale training environments, when a job fails, the system detects the fault, performs isolation, and initiates recovery procedures. Once resolved, the training framework resumes from the most recently saved checkpoint. Common fault recovery mechanisms based on checkpointing are categorized into three levels, which is shown in Fig.\ref{threelevel}:

\begin{figure}[t]
    \centering
    \includegraphics[scale=0.6]{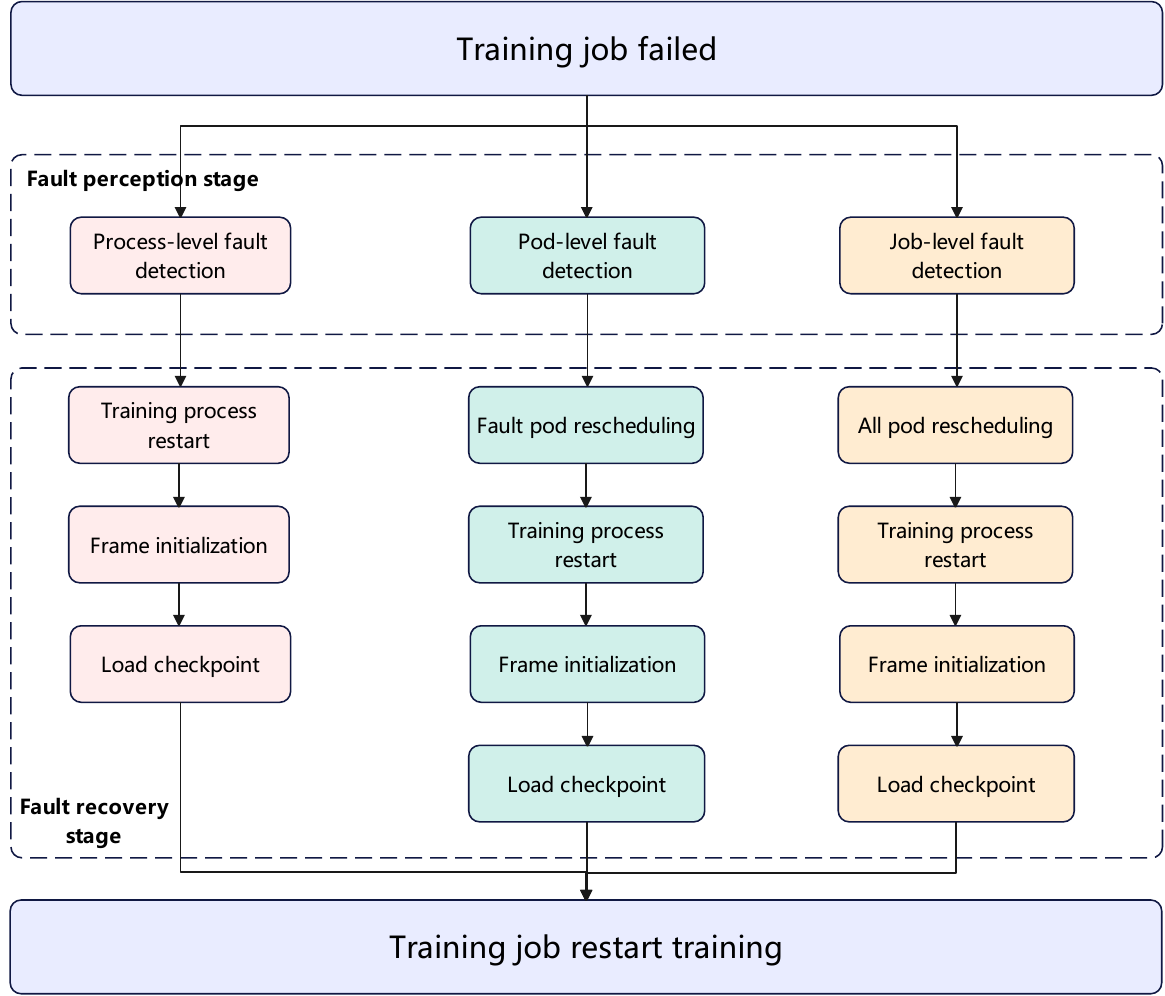}
    \caption{Three-level Strategy for Fault Tolerance.}
    \label{threelevel}
\end{figure}

\begin{itemize}
\item Process-Level Fault Handling: This category involves failures caused by software errors or training process crashes, detected by the training framework. Recovery simply requires restarting the affected processes without rescheduling any nodes within the job.
\item Pod-Level Fault Handling: These failures stem from hardware issues (e.g., GPU errors, node failures) or parameter network problems, detected by the Kubernetes (K8s) scheduler. Recovery involves rescheduling only the faulty Pod(s) and restarting the training processes, minimizing disruption to the overall job.
\item Job-Level Fault Handling: This level addresses failures due to unknown causes that result in abnormal training termination (indicated by specific exit codes), detected by the K8s scheduler. Recovery requires rescheduling all nodes associated with the job and restarting the training processes.
\end{itemize}

\subsection{Pretraining Performance Modeling Techniques}
White-box simulation methods, like Galvatron~\cite{liu2025galvatron} and AutoPlan~\cite{li2024automatically}, addressed the challenge of identifying optimal parallel strategies by decomposing and integrating computational, communication, and memory modules. In particular, Galvatron automates the identification of the most efficient hybrid parallelization strategies, thereby overcoming the complexity of selecting an optimal parallel configuration. Its architecture comprises a performance profiler for hardware and model analysis, a search engine that utilizes decision trees and dynamic programming for strategy optimization, and a runtime system for executing the selected strategies. AutoPlan introduces a method for automated parallel strategy planning in large-scale language model pretraining. It decomposes training time into computation and communication,  establishing a training duration simulation model. The approach formulates an automatic parallelization algorithm based on model and hardware characteristics, planning parallel strategies for maximum throughput. By incorporating additional variables and pruning the search space, the algorithm can more accurately evaluate different parallel strategies and identify the optimal strategy combination.

Trace-driven methods rely on runtime traces to construct fine-grained execution graphs, replaying the actual execution process for performance estimation, with Lumos~\cite{liang2025lumos} being a representative example. Lumos is a trace-based performance modeling and estimation toolkit designed to accurately capture and predict the execution behavior of modern LLMs. It leverages PyTorch Kineto to collect trace information and construct execution graphs, offering a user-friendly interface that requires only a few lines of hook code for performance analysis. The toolkit supports optimization exploration across various configurations and model types, including distributed inference and SSD-based inference. By simulating existing execution graphs, Lumos can predict performance under new configurations, thereby providing developers with valuable optimization suggestions.

Although these methods exhibit respective advantages in strategy search efficiency or iteration time estimation, they still suffer from inadequate consideration of optimization features and a lack of fault tolerance modeling. To address these limitations, we propose MoFa to characterize pretraining behavior and design a MoFa tuning system, which provides guidance for practical deployment and tuning of LLM pretraining.

%% file: mofadesign.tex
\section{MoFa Design}
This section elaborates on the design of MoFa. We first define and describe the critical metrics used in the design of the cost model, followed by the base cost model of the pretraining system, denoted as $\mathrm{CostModel}_\mathrm{Base}$. We then introduce the multi-dimensional optimization-aware cost model $\mathrm{CostModel}_\mathrm{Mo}$ and the fault tolerance cost model $\mathrm{CostModel}_\mathrm{Fa}$ in MoFa. 


\subsection{Definition of Key Factors}

In large-scale model pretraining systems, numerous factors influence pretraining efficiency, such as model architecture, hardware parameters, and parallelization strategies. To establish a unified modeling framework, it is essential to abstract each category of features into semantically well-defined quantitative parameters. This abstraction enables systematic performance analysis and prediction across different pretraining configurations. The key metrics affecting pretraining efficiency in the system are summarized in Table \ref{tab:definition}.


\input{Table/3/notation}

\subsection{Basic Performance Cost Model}

\begin{figure}[t]
\centering
    \includegraphics[trim=40mm 60mm 40mm 50mm, clip, scale=0.64]{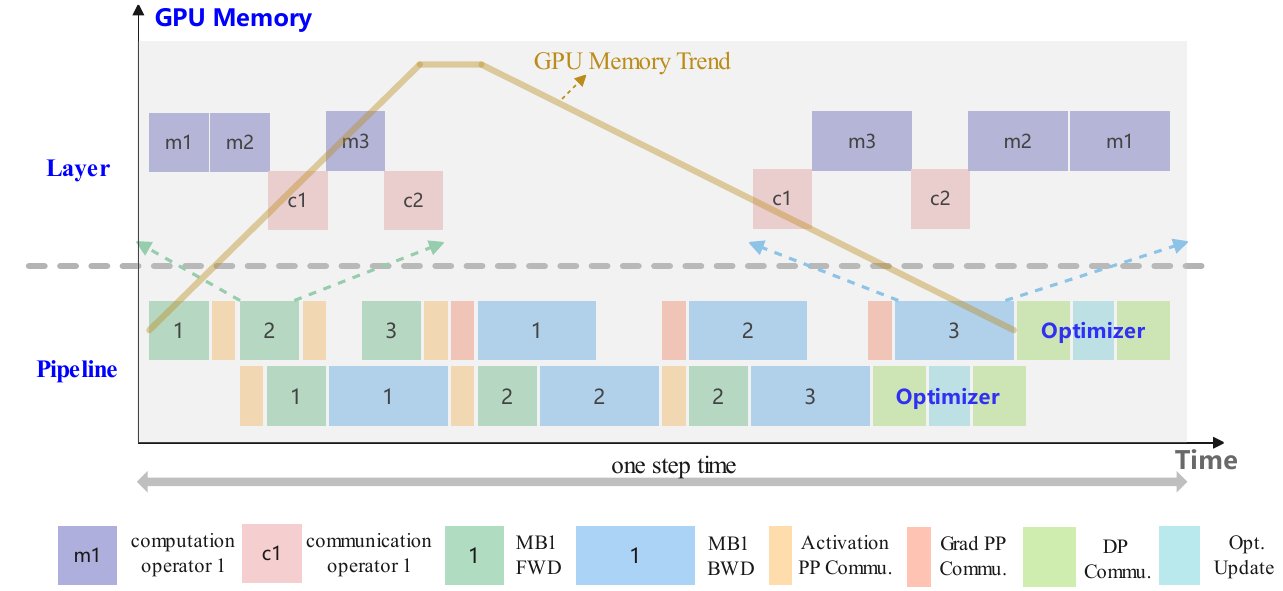}
    \caption{Timing diagram of one pretraining step.}
    \label{one-step}
\end{figure}

This section introduces the basic cost model for the pretraining system, denoted as $\mathrm{CostModel}_\mathrm{base}$, which mathematically characterizes the fundamental components and workflows of large-scale pretraining systems. Fig.\ref{one-step} presents a timing diagram of a typical pretraining step. Based on this workflow, $\mathrm{CostModel}_\mathrm{base}$ incorporates six key components: model architecture decomposition, operator and communication latency modeling, layer-level latency modeling, pipeline-stage latency modeling, optimizer latency modeling, and memory consumption analysis.

\textbf{Model Architecture Decomposition}: Modern LLMs predominantly adopt decoder-only architectures, with the most prevalent being Dense, like the LLaMA series, and MoE structures, like DeepSeek-V3. We decompose model architectures hierarchically into three levels. The first two levels remain identical for both Dense and MoE variants, while differentiation occurs at the third level within the Multi-Layer Perceptron (MLP) component. Table~\ref{three-level} summarizes common architectural configurations, where the attention module is represented by Multi-Head Attention (MHA). In practice, attention implementations may vary, e.g., MHA, Grouped-Query Attention (GQA)~\cite{ainslie2023gqa}, Multi-Query Attention (MQA)~\cite{ainslie2023gqa},, and Multi-Head Latent Attention (MLA)~\cite{liu2024deepseek},, significantly impacting both computational Floating-point operations (FLOPs) and stored activation sizes.

\textbf{Computation and Communication Latency Modeling}: Computational operators and communication operations constitute the foundational elements of distributed pretraining performance. We employ profiling-based measurements to model computational throughput $P^{\mathrm{comp}}$ and communication bandwidth $B^{\mathrm{commu}}$, prioritizing accuracy over ideal assumptions. The latency models are formulated as:
\begin{align}
\label{comp}
    T_i^{\mathrm{comp}} &= S_i^{\mathrm{comp}} / P_i^{\mathrm{comp}} \\
\label{commu1}
    T_j^{\mathrm{commu}} &= S_j^{\mathrm{commu}} / B_j^{\mathrm{commu}}
\end{align}
where $S_i^{\mathrm{comp}}$ denotes the computation of the $i$-th operator, $P_i^{\mathrm{comp}}$ represents profiled throughput, $S_j^{\mathrm{commu}}$ indicates the $j$-th communication, and $B_j^{\mathrm{commu}}$ reflects measured bandwidth. While the profiling of computation operators can be conducted with limited hardware resources, accurate communication benchmarking at ultra-scale remains challenging. To address this, we measure the bandwidth on small-scale setups and apply empirical scaling factors on large-scale clusters:
\begin{align}
\label{commu2}
    T_j^{\mathrm{commu}} &= \frac{S_j^{\mathrm{commu}}}{\beta_j B_j^{\mathrm{commu}}}
\end{align}
where $\beta_j$ denotes bandwidth decay factor from small-scale clusters to large-scale clusters.

\input{Table/3/modelarchitecture}

\textbf{Memory Cost Model}: Memory consumption of LLM pretraining on GPUs comprises static and dynamic components. Static memory includes weight parameters, gradients, and optimizer states, modeled as,
\begin{align}
\label{staticmem}
M^{\mathrm{sta}} = \underbrace{D_t^{\mathrm{para}} v l \sum_i^{m_n} S_i^{\mathrm{para}}}_{\mathrm{Parameters}} + \underbrace{D_t^{\mathrm{grad}} v l \sum_i^{m_n} S_i^{\mathrm{para}}}_{\mathrm{Gradients}} + \underbrace{4 D_t^{\mathrm{opt}} v l \sum_i^{m_n} S_i^{\mathrm{para}}}_{\mathrm{Optimizer\ States}}
\end{align}
where $S_i^{\mathrm{para}}$ represents the size of $i$-th weight on per GPU, and $D_t^{\mathrm{para}}$, $D_t^{\mathrm{grad}}$, $D_t^{\mathrm{opt}}$ respectively denote the data types of weight, grad and optimizer states.

Dynamic memory primarily arises from activation storage during forward propagation for backward pass computation. Considering the PP scheduling pattern, as shown in Fig.\ref{one-step}, which follows warmup-steady-cooldown phases, activation memory peaks during warmup completion:
\begin{align}
\label{actmem}
M^{\mathrm{act}} = \left(v p + p - 2 r_{pp} - 1 \right) \sum_i^{m_n} M_i^{\mathrm{act}}
\end{align}
where $M_i^{\mathrm{act}}$ denotes the stored activation size of the $i$-th module and $r_{pp}$ the pipeline stage index. Obviously, the first pipeline stage experiences maximum memory pressure, e.g. $\left(v p + p - 1 \right) \sum_i^{m_n} M_i^{\mathrm{act}}$.

Therefore, the peak memory consumption is 
\begin{align}
\label{allmem}
M^{\mathrm{peak}} = M^{\mathrm{sta}} + M^{\mathrm{act}}
\end{align}

\textbf{Layer-level Cost Model}: This component characterizes forward (FWD) and backward (BWD) latency per transformer layer:
\begin{align}
\label{baseFWD}
T^{\mathrm{FWD}} &= \sum\nolimits_{i=1}^{m_n} T_i^{\mathrm{comp}} + \sum\nolimits_{j=1}^{c_n} T_j^{\mathrm{commu}} \\
\label{baseBWD}
T^{\mathrm{BWD}} &= \sum\nolimits_{i=1}^{m_n} T_i^{\mathrm{comp_{BWD}}} + \sum\nolimits_{j=1}^{c_n} T_j^{\mathrm{commu}}
\end{align}
where $T_i^{\mathrm{comp}}$ and $T_i^{\mathrm{comp_{BWD}}}$ represent the forward and backward latency of the $i$-th computation, respectively, and $T_j^{\mathrm{commu}}$ denotes the latency of the $j$-th communication.

\textbf{Pipeline-level Cost Model}: We adopt the interleaved One-Forward-One-Backward (1F1B) scheduling strategy with phase-wise modeling~\cite{yuan2024accelerating}:
\begin{align}
\label{ppwarmup}
T^{\mathrm{Warmup}} &= p(T^{\mathrm{Embed}} + l T^{\mathrm{FWD}} + T^{\mathrm{PP}}) + (v p - p - 1)(l T^{\mathrm{FWD}} + T^{\mathrm{PP}}) \\
\label{ppsteady}
T^{\mathrm{Steady}} &= 
\begin{aligned}[t]
 &p(l T^{\mathrm{FWD}} + T^{\mathrm{Head}} + T^{\mathrm{Head_{BWD}}} + l T^{\mathrm{BWD}}) \\
 &\quad + (m_b - p)(v l T^{\mathrm{FWD}} + T^{\mathrm{Head}} + T^{\mathrm{Head_{BWD}}} + l T^{\mathrm{BWD}}) \\
 &\quad + (4 m_b v - 2 m_b + 2 p - 2) T^{\mathrm{PP}}
\end{aligned} \\
\label{ppcooldown}
T^{\mathrm{Cooldown}} &= p(T^{\mathrm{Embed_{BWD}}} + l T^{\mathrm{BWD}} + T^{\mathrm{PP}}) + (v p - p - 1)(l T^{\mathrm{BWD}} + T^{\mathrm{PP}}) \\
\label{pipeline}
T^{\mathrm{Pipeline}} &= T^{\mathrm{Warmup}} + T^{\mathrm{Steady}} + T^{\mathrm{Cooldown}}
\end{align}
where $T^{\mathrm{Embed}}$, $T^{\mathrm{Embed_{BWD}}}$, $T^{\mathrm{Head}}$ and $T^{\mathrm{Head_{BWD}}}$ denote the latencies of embedding and head, respectively.

\textbf{Optimizer-level Modeling}: The optimizer schedule combines DP communication and parameter updates:
\begin{align}
\label{optimizer}
\nonumber
T^{\mathrm{Opt}} &= T^{\mathrm{DP}} + T^{\mathrm{update}} \\
                &= \frac{D_t^{\mathrm{grad}} v l \sum_i^{m_n} S_i^{\mathrm{para}}}{B^{\mathrm{DP}}} + \frac{v l \sum_i^{m_n} S_i^{\mathrm{para}}}{P^{\mathrm{Opt}}}
\end{align}
where $P^{\mathrm{Opt}}$ represents the computational throughput of the optimizer update.

The latency and FLOPS of per pretraining step as shown in Fig.\ref{one-step} are consequently modeled as:
\begin{align}
\label{stepbase}
    T^\mathrm{step} = T^\mathrm{Pipeline}+T^\mathrm{Opt}
\end{align}
\begin{align}
\label{tflops}
    \mathrm{TFLOPS} = \frac{(\mathrm{FLOPs}_\mathrm{Embed}+ \mathrm{FLOPs}_\mathrm{Head} + L\sum_i^{m_n} \mathrm{FLOPs}_i)/10^{12}}{T^\mathrm{step}}
\end{align}

\subsection{Multi-Dimensional Optimizations-Aware Performance Cost Model}
In practical large-scale model pretraining, parallel strategies and optimization features simultaneously impact pretraining efficiency. Different optimization features operate at various levels of the foundational modeling, influencing the optimal selection of parallel strategies. To unify the impact of optimization features on the performance of the pretraining system, we introduce multi-dimensional optimization-aware modeling atop the foundational cost model $\mathrm{CostModel}_\mathrm{Mo}$ , enabling more precise and comprehensive modeling of pretraining system performance. There $\mathrm{CostModel}_\mathrm{Mo}$ is a unified modeling framework that incorporates mainstream optimization features based on \eqref{stepbase}, covering Layer-level Optimization (LO), Pipeline-level Optimization (PO), Optimizer-level Optimization (OO), and Memory-level Optimization (MO), expressed as follows
\begin{align}
\label{costmodelmofa}
    \mathrm{CostModel}_\mathrm{Mo} = \mathrm{CostModel}_\mathrm{base}\left(LO,PO,OO,MO\right)
\end{align}
For each category of optimization features, we design a cost model to describe their impact on the performance of the pretraining system.

\subsubsection{Layer-level Optimization Modeling}
Layer-level optimization refers to optimizations affecting \eqref{baseFWD} and \eqref{baseBWD}, including layer-level communication bandwidth enhancement, layer-level computational throughput improvement, and layer-level computation-communication overlap design. Layer-level communication bandwidth enhancement optimizes communication algorithms to increase bandwidth, while layer-level computational throughput improvement optimizes computation implementation algorithms to enhance computational power. To uniformly express the impact of these two aspects on performance, we model their effects by introducing scaling coefficients, $\lambda_i^\mathrm{comp}$ and and $\lambda_j^\mathrm{commu}$, as follows
\begin{align}
\label{suanzi-up}
P_i^{\mathrm{comp}}=\lambda_i^\mathrm{comp}P_i^\mathrm{comp}
\end{align}
\begin{align}
\label{tongxin-up}
B_j^{\mathrm{commu}}=\lambda_j^\mathrm{commu}B_j^\mathrm{commu}
\end{align}
To explore the efficiency of computation, theoretical cost models, e.g., roofline models~\cite{imai2024predicting,wang2025step}, have been widly developed. roofline model is defined as
\begin{align}
    P_i^{\mathrm{peak}} = \min(\frac{S_c}{S_m}B^\mathrm{HBM},P^\mathrm{GPU})
\end{align}
where $S_c$ and $S_m$ are FLOPs and memory traffic number at sofrware, respectively. $B^\mathrm{HBM}$ denotes as the memory bandwidth. Roofline model provides an upper bound for the computational power. 

Substitute \eqref{suanzi-up} and \eqref{tongxin-up} into \eqref{comp} and \eqref{commu2}, respectively, to characterize the performance impact of computational power improvement and bandwidth enhancement.

Layer-level communication-computation overlap design includes TP overlap, CP overlap, and EP overlap. We model each strategy according to its design principles.

\textbf{TP overlap} typically involves pipelining adjacent matrix multiplication operators and TP communication by splitting them into multiple stages for overlapping, e.g., overlapping the QKV generation matrix multiplication operation with TP all-gather (ag). Under this design, the effective exposed time after overlapping TP communication and computation is modeled as follows:
\begin{align}
\label{tp-overlap-costmodel}
T_i^{\mathrm{comp}} + T_j^{\mathrm{TP}} = \frac{1}{s_n} \min \Biggl(T_i^{\mathrm{comp}},T_j^{\mathrm{TP}} \Biggr)+
\max \Biggl( \alpha_i ^\mathrm{comp}T_i^{\mathrm{comp}}, \beta_j ^\mathrm{TP}T_j^\mathrm{TP} \Biggr)
\end{align}
where $s_n$ is the number of split stages, $T_j^{\mathrm{TP}}$ and $T_i^{\mathrm{comp}}$ represent the $j$-th TP communication and the $i$-th computation module participating in the overlap, respectively, and $\alpha_i ^\mathrm{comp}$ and $\beta_j ^\mathrm{TP}$ are the time increase coefficients for the $i$-th computation and $j$-th TP communication during overlap, typically due to hardware resource contention.

\textbf{CP overlap} is usually designed using ring-attention, where the $i$-th attention computation overlaps with the peer-to-peer communication for the $(i+1)$-th sequence batch. Under this design, the effective exposed time after CP overlap is modeled as follows:
\begin{align}
\label{cp-overlap-costmodel}
T_i^{\mathrm{attention}} + T_j^{\mathrm{CP}} = \frac{1}{c} \min \Biggl(T_i^{\mathrm{attention}},T_j^{\mathrm{CP}} \Biggr)+
\max \Biggl( \alpha_i ^\mathrm{attention}T_i^{\mathrm{attention}}, \beta_j ^\mathrm{CP}T_j^\mathrm{CP} \Biggr)
\end{align}
where $T_i^{\mathrm{attention}}$ and $T_j^{\mathrm{CP}}$ denote the $i$-th attention computation and the $j$-th CP communication in ring-attention, respectively, and $\alpha_i ^\mathrm{attention}$ and $\beta_j ^\mathrm{CP}$ are the time increase coefficients for the $i$-th computation and $j$-th CP communication during overlap.

\textbf{EP overlap} typically employs a design similar to overlapping forward and backward passes. Specifically, within a layer, the computation of the forward pass is overlapped with the EP communication of the backward pass, and the computation of the backward pass is overlapped with the EP communication of the forward pass. Under this design, the effective exposed time after EP overlap is uniformly modeled as follows:
\begin{align}
\label{ep-overlap-costmodel}
    \sum\nolimits_{i=1}^{m_n}T_i^{\mathrm{comp}} + \sum\nolimits_{j=1}T_j^{\mathrm{EP}} = \max \left(\alpha _i^{\mathrm{comp}}\sum\nolimits_{i=1}^{m_n}T_i^{\mathrm{comp}}, \beta _j^{\mathrm{EP}}\sum\nolimits_{j=1}^{c_n}T_j^{\mathrm{EP}}\right)
\end{align}
where $T_j^{\mathrm{EP}}$ and $T_i^{\mathrm{comp}}$ represent the $j$-th EP communication and the $i$-th computation module participating in the overlap, respectively, and $\alpha_i ^\mathrm{comp}$ and $\beta_j ^\mathrm{EP}$ are the time increase coefficients for the $i$-th computation and $j$-th EP communication during overlap.

Applying \eqref{suanzi-up} to \eqref{ep-overlap-costmodel} to \eqref{baseFWD} and \eqref{baseBWD} yields the layer-level optimization enhanced cost model, characterizing the impact of layer-level optimizations on the per-layer time cost.

\subsubsection{Pipeline-level Optimization Modeling}
Pipeline-level optimization refers to optimizations affecting \eqref{pipeline}. Pipeline-level optimizations include layer-level optimizations and optimizations related to $T^\mathrm{PP}$. Layer-level optimizations were introduced in the previous subsection, and incorporating them into \eqref{pipeline} captures their performance impact on the pipeline. This subsection focuses on modeling optimizations related to $T^\mathrm{PP}$, primarily PP communication bandwidth enhancement and PP overlap design. The modeling for PP communication bandwidth enhancement is consistent with \eqref{commu2}.

The PP overlap design utilizes the forward or backward process of the next micro-batch (MB) to overlap the adjacent PP communication process, primarily affecting the steady phase in 1F1B. We model this PP overlap as follows:
\begin{align}
\label{PPOverlap}
    T^{\mathrm{PP}} = \max \Biggl(0, \beta_j^{\mathrm{PP}}T^{\mathrm{PP}}-\alpha_k\sum\nolimits_{k=1}^{l}T_k^\mathrm{onelayer} \Biggr)
\end{align}
where $T_k^\mathrm{onelayer}$ represents the time cost of the $k$-th layer. Substitute \eqref{baseFWD} when the forward process overlaps PP communication, and \eqref{baseBWD} when the backward process overlaps PP communication. $\beta_j^{\mathrm{PP}}$ and $\alpha_k$ are the time increase coefficients for the $j$-th PP module and the $k$-th layer computation, respectively. Incorporating \eqref{PPOverlap} into \eqref{ppsteady} characterizes the impact of PP overlap on pipeline performance.

\subsubsection{Optimizer-level Optimization Modeling}
Optimizer-level Optimization refers to optimizations affecting \eqref{optimizer}, primarily divided into two parts: DP communication optimization and parameter update acceleration.
DP communication optimization includes two types: DP communication bandwidth enhancement and DP overlap design. The modeling for DP communication bandwidth enhancement is consistent with \eqref{commu2}. The DP  overlap design typically employs the pipeline computation process to overlap DP communication. Specifically, the cooldown phase of the pipeline overlaps the reduce-scatter (RS) of gradients, i.e., the backward process of the $i$-th virtual pipeline overlaps the gradient RS of the $(i+1)$-th virtual pipeline. The warmup phase overlaps the AG of parameters, i.e., the forward process of the $i$-th virtual pipeline overlaps the parameter AG of the $(i+1)$-th virtual pipeline. We model this design as follows:
\begin{align}
\label{dpoverlap}
    T^{\mathrm{DP}} =  T_{\mathrm{chunk}_1}^{\mathrm{DP}_{\mathrm{rs}}} + T_{\mathrm{chunk}_1}^{\mathrm{DP}_{\mathrm{ag}}} + &\max \Biggl(\alpha ^{\mathrm{DP}_{\mathrm{rs}}}\sum\nolimits_{i=2}^{v}T_{\mathrm{chunk}_i}^{\mathrm{DP}_{\mathrm{rs}}}, \beta ^{\mathrm{BWD}}pl(v-1)T^{\mathrm{BWD}}\Biggr) +
    \\ \nonumber
    & \max \Biggl(\alpha ^{\mathrm{DP}_{\mathrm{ag}}}\sum\nolimits_{j=2}^{v}T_{\mathrm{chunk}_j}^{\mathrm{DP}_{\mathrm{ag}}}, \beta ^{\mathrm{FWD}}pl(v-1)T^{\mathrm{FWD}}\Biggr)
\end{align}
where $T_{\mathrm{chunk}_i}^{\mathrm{DP}_{\mathrm{rs}}}$ and $T_{\mathrm{chunk}_j}^{\mathrm{DP}_{\mathrm{ag}}}$ are the time costs for the RS of the $i$-th virtual pipeline and the AG of the $j$-th virtual pipeline, respectively. $\alpha ^{\mathrm{DP}_{\mathrm{rs}}}$ and $\beta ^\mathrm{BWD}$ are the time increase coefficients for DP RS and backward propagation, respectively, and $\alpha ^{\mathrm{DP}_{\mathrm{ag}}}$ and $\beta ^\mathrm{FWD}$ are the time increase coefficients for DP AG and forward propagation, respectively.

Parameter update acceleration belongs to operator optimization, primarily optimizing the parameter update algorithm to enhance computational power. Its modeling is consistent with \eqref{suanzi-up}.

Substituting \eqref{dpoverlap} and \eqref{suanzi-up} into \eqref{optimizer} describes the impact of optimizer-level optimizations on the optimizer time cost.

\subsubsection{Memory-level Optimization Modeling}
As model pretraining scales continue to expand, the demand for memory has grown increasingly, making memory optimization a crucial technique for pretraining acceleration. Current mainstream distributed pretraining frameworks commonly support various memory saving techniques, which can be categorized into two types: optimizer state memory optimization and activation memory optimization. To quantify the actual effects of such optimizations, we establish cost models for memory usage and pretraining performance for the aforementioned strategies.

\textbf{Optimizer State Memory Optimization}: Optimization strategies include distributed optimizer and CPU optimizer, both of which can significantly reduce optimizer state memory usage but have different performance impacts. The distributed optimizer partitions the optimizer states across the DP group, reducing the optimizer memory footprint per DP rank. Each DP rank is responsible for maintaining and updating 1/DP of the parameters, thereby also reducing the parameter update time by (DP-1)/DP. Based on this, the memory and performance modeling under the distributed optimizer feature is as follows:
\begin{align}
\label{distributed-optimizer-memory}
    & M^{\mathrm{optimizer}} = \frac{4D_t^\mathrm{opt}vl\sum_i^{m_n}S_i^\mathrm{para}}{d} \\
\label{distributed-optimizer-performance}
    & T^\mathrm{update} = \frac{T^\mathrm{update}}{d}
\end{align}

The CPU optimizer offloads optimizer states to the CPU, performs parameter updates on the CPU, and then transfers the updated parameters back to the GPU. This can reduce nearly 100\% of the optimizer memory footprint but introduces significant H2D and D2H transfer times and CPU parameter update time. We model the memory and performance under the CPU optimizer feature as follows:
\begin{align}
\label{cpuadam-mem}
& M^{\mathrm{optimizer}} = \max \Biggl(0, 4D_t^\mathrm{opt}vl\sum_i^{m_n}S_i^\mathrm{para} - M^{\mathrm{cpu}}\Biggr) \\
\label{cpuadam-performance}
& T^\mathrm{update} = \frac{\sum_i^{m_n}S_i^\mathrm{para}}{F^\mathrm{cpu}}+\frac{D_t^\mathrm{grad}vl\sum_i^{m_n}S_i^\mathrm{para}}{B^\mathrm{h2d}}+\frac{D_t^\mathrm{para}vl\sum_i^{m_n}S_i^\mathrm{para}}{B^\mathrm{d2h}}
\end{align}

\textbf{Activation Memory Optimization}: Activation memory saving typically employs two types of optimizations: activation recomputation or activation offloading. Similarly, these two optimizations have different impacts on pretraining performance.

Activation Recomputation: Activation recomputation includes two types: selective recomputation and full recomputation. Selective recomputation only recomputes the activation values of the attention part, preserving other activation values. During BWD, it recomputes the generations of the Q matrix, K matrix, and V matrix to produce the inputs for attention module and recomputes the attention module to generate intermediate activation values for attention. We model this optimization as follows:
\begin{align}
    & M^{\mathrm{act}} = \left(vp+p-2r_{pp}-1 \right)\times\Biggl(\sum_i^{m_n}M_i^\mathrm{act}-M^\mathrm{Attention}\Biggr)\\
    & T^\mathrm{BWD} = T^\mathrm{BWD}+T^\mathrm{QKV}+T^\mathrm{Attention}
\end{align}
where $M^\mathrm{Attention}$ is the intermediate activation of the Attention module, and $T^\mathrm{QKV}$ and $T^\mathrm{Attention}$ are the computation times for the Q matrix, K matrix, V matrix, and the attention module, respectively.

Full recomputation involves recomputing an entire transformer layer, retaining only the layer's input, and recomputing the forward pass of the layer during backward propagation to regenerate the activation values, which can significantly save activation memory. Based on the characteristics of the two recomputation designs, we model the memory and performance as follows:
\begin{align}
\label{full-mem}
& M^{\mathrm{act}} = \left(vp+p-2r_{pp}-1 \right)M^\mathrm{input}\\
\label{full-performance}
& T^\mathrm{BWD} = T^\mathrm{BWD} + T^\mathrm{FWD}
\end{align}
where $M^\mathrm{input}$ is the memory size occupied by the input of one layer.

Activation Offloading: Mainstream activation offloading strategies typically employ asynchronous activation offloading. The principle is to asynchronously offload the activation values of each layer during the forward propagation of the next layer and prefetch the activation values of each layer during the backward propagation of the previous layer. This approach can reduce a significant amount of activation memory usage while introducing less computational overhead. Based on this design characteristic, we model the asynchronous activation offloading in terms of memory and performance as follows:
\begin{align}
\label{offload-mem}
& M^{\mathrm{act}} = \sum_i^{m_n}M_i^\mathrm{act}\\
\label{offload-fwd}
& T^\mathrm{FWD} = \max \Biggl( \alpha^\mathrm{offload} \frac{M^{\mathrm{act}}}{B^\mathrm{D2H}}, \beta^\mathrm{offload} T^\mathrm{FWD} \Biggr) \\
\label{offload-bwd}
& T^\mathrm{BWD} = \max \Biggl( \alpha^\mathrm{fetch} \frac{M^{\mathrm{act}}}{B^\mathrm{H2D}}, \beta^\mathrm{fetch} T^\mathrm{BWD} \Biggr)
\end{align}
where $\alpha^\mathrm{offload}$ and $\beta^\mathrm{offload}$ are the time increase coefficients for D2H transfer and computation during offloading, respectively, and $\alpha^\mathrm{fetch}$ and $\beta^\mathrm{fetch}$ are the time increase coefficients for H2D transfer and computation during activation fetching, respectively.

Substituting the selected memory optimization strategy into the affected foundational performance model allows for assessing the impact of the respective memory optimization on memory usage and performance, thereby effectively evaluating pretraining efficiency.

\subsection{Fault Tolerance Performance Cost Model}
In practical pretraining systems, beyond the standard pretraining duration, two critical factors contribute to the E2E runtime of a pretraining job: the time spent on fault recovery due to failures and the time allocated for checkpoint saves during pretraining. Therefore, the total E2E pretraining duration of a job (excluding job queuing time on the cluster, as the pretraining job is the high-priority task) comprises the pretraining effective time, the first pretraining initialization time, the fault recovery time, and the checkpoint saving time. Moreover, the effective pretraining duration consists solely of pretraining effective time, while the pretraining interruption duration is composed of first pretraining initialization time, fault recovery time, and checkpoint saving time.

Inspired by~\cite{DBLP:conf/hpca/KokolisKHKMMDSS25}, the effective pretraining time ratio (ETTR) is defined as
\begin{align}
\label{ettr}
    ETTR = \frac{T^\mathrm{tr}}{T^\mathrm{tr}+T^\mathrm{in}}
\end{align}
Where $T^\mathrm{tr}$ denotes the effective pretraining duration, and $T^\mathrm{in}$ denotes the pretraining interruption duration. The effective pretraining duration is defined as
\begin{align}
\label{en}
    T^\mathrm{tr}= S\times T^\mathrm{step}
\end{align}
Where $S$ denotes the number of total pretraining step.

To calculate the pretraining interruption duration, we first need to determine the number of failures that occurred during the pretraining cycle. The number of failures during the pretraining phase is defined as
\begin{align}
\label{en}
    F_f=N^\mathrm{nodes}r_f\left(T^\mathrm{tr}+T^\mathrm{in}\right)
\end{align}
Where $N^\mathrm{nodes}$ denotes the total number of nodes, which is equal to $G_n/N$, and $r_f$ denotes the number of failures a node experiences per day. Then the pretraining interruption duration is defined as
\begin{align}
\label{en}
    T^\mathrm{in}= u_0 + F_fu_b + F_f\frac{I^\mathrm{ckpt}T^\mathrm{step}}{2} + \lceil \frac{S}{I^\mathrm{ckpt}} \rceil T^\mathrm{save}
\end{align}
Where $u_0$ denotes the first pretraining initialization time, $u_b$ denotes the mean time to repair per failure, $I^\mathrm{ckpt}$ denotes the checkpoint saving interval, and $T^\mathrm{save}$ denotes the single checkpoint save time. From \eqref{ettr}$\sim$\eqref{en}, the definition of ETTR is ultimately obtained as
\begin{align}
    \mathbb{E}(ETTR)=\frac{1-N^\mathrm{nodes}r_f\left(u_b+\frac{I^\mathrm{ckpt}T^\mathrm{step}}{2}\right)}{1+\frac{T^\mathrm{save}}{I^\mathrm{ckpt}T^\mathrm{step}}}
\end{align}
The calculation of $u_b$ has undergone significant changes with the evolution of fault-tolerant technologies. Previously, after the pretraining job failed, regardless of the specific fault type, the job would restart once, meaning all nodes associated with the job required rescheduling. With the advancement of fault tolerance technology, the current art-of-state fault recovery methods are broadly categorized into three levels: process-level faults recovery, POD-level faults recovery, and job-level faults recovery, as shown in Fig.\ref{threelevel}. The recovery times from shortest to longest are: process-level fault recovery, POD-level fault recovery, and job-level fault recovery. Under this fault recovery model, the mean time to repair per failure $u_b$ is defined as
\begin{align}
\label{en}
    u_b = \alpha u_{bc} + \beta u_{bp} + \gamma u_{bj}
\end{align}
Where $\alpha$, $\beta$, $\gamma$ denote the probability for process-level failures, POD-level failures, and job-level failures, $u_{bc}$, $u_{bp}$, $u_{bj}$ denote the process-level fault recovery duration, the POD fault recovery duration, and the job-level fault recovery duration. By observing historical data from thousands of GPU cluster, the failure probability ratio of $\alpha$, $\beta$, and $\gamma$ is approximately 3:6:1. The average time to restore the single fault $u_{bc}$, $u_{bp}$, $u_{bj}$ is 141s, 262s, and 307s.

%% file: Table/3/notation.tex
\begin{table}[t]
\centering
\caption{Metrics Affecting Training Efficiency}
\label{tab:definition} 
\renewcommand{\arraystretch}{1.2}
\scalebox{0.85}{
\begin{tabular}{ccc} \hline
                            & Notation           & Instructions             \\ \hline
\multirow{7}{*}{Model Architecture} & $L$    & transformer layers         \\
                            & $h$   & hidden dimension size       \\
                            & $s$    & sequence length          \\
                            & $g_d$  & dense intermediate size         \\
                            & $g_e$  & moe intermediate size     \\
                            & $a$    & attention heads         \\
                            & $V$    & vocabulary size       \\ 
                            \hline
\multirow{9}{*}{Parallelism Configuration} & $t$    & tensor parallel size             \\
                            & $p$    & pipeline parallel size              \\
                            & $c$    & context parallel size            \\
                            & $d$    & data parallel size            \\
                            & $e$    & expert parallel size            \\
                            & $m_{bs}$ & micro batch size  \\
                            & $g_{bs}$ & global batch size  \\
                            & $v$    & pipeline stage per GPU \\
                            & $l$    & layer per stage \\
                            & $m_n$    & module number per layer \\
                            \hline
\multirow{9}{*}{Hardware Configuration}   & $B^\mathrm{H2D}$ & host to device (H2D) bandwidth (GB/s)\\
                            & $B^\mathrm{D2H}$ & device to host (D2H) bandwidth (GB/s) \\ 
                            & $B^\mathrm{DL}$ & disk load bandwidth (GB/s)\\ 
                            & $B^\mathrm{DW}$ & disk write bandwidth (GB/s)\\
                            & $M^\mathrm{CPU}$ & Central Processing Unit (CPU) memory (GB) \\
                            & $F^\mathrm{CPU}$ & CPU frequency (GHz)\\
                            & $P^\mathrm{GPU}$ & GPU specific computing power (TFLOPS) \\
                            & $M^\mathrm{GPU}$ & GPU memory (GB) \\
                            & $N$ & GPU number per node \\
                            \hline
\end{tabular}}
\end{table}

%% file: Table/3/modelarchitecture.tex
\begin{table}[t]
\centering
\caption{Three-level architecture of LLM.}
\label{three-level}
\renewcommand{\arraystretch}{1.2}
\scalebox{0.9}{
\begin{tabular}{|c|c|cc|cc|cc|}
\hline
\textbf{First Level} & \textbf{Second Level}               & \multicolumn{2}{c|}{\textbf{Third Level}}        & \multicolumn{2}{c|}{\textbf{FLOPs}}          & \multicolumn{2}{c|}{\textbf{Activation size}} \\ \hline
Dense, MoE  & Dense, MoE                 & \multicolumn{1}{c|}{Dense}    & MoE     & \multicolumn{1}{c|}{Dense}  & MoE    & \multicolumn{1}{c|}{Dense}   & MoE   \\ \hline
Embedding   & -                          & \multicolumn{2}{c|}{-}                  & \multicolumn{1}{c|}{$bsh$}    & $bsh$    & \multicolumn{1}{c|}{$2bsh$}    & $2bsh$  \\ \hline
\multirow{11}{*}{\begin{tabular}[c]{@{}c@{}}Transformer Layer \\ $\times L$\end{tabular}} &
  norm &
  \multicolumn{2}{c|}{-} &
  \multicolumn{1}{c|}{$bsh$} &
  $bsh$ &
  \multicolumn{1}{c|}{$2bsh$} &
  $2bsh$ \\ \cline{2-8} 
            & \multirow{5}{*}{attention} & \multicolumn{2}{c|}{Q,K,V}              & \multicolumn{1}{c|}{$6bsh^2$}  & $6bsh^2$  & \multicolumn{1}{c|}{$2bsh$}    & $2bsh$  \\ \cline{3-8} 
            &                            & \multicolumn{2}{c|}{attention map}      & \multicolumn{1}{c|}{$2bs^2h$}  & $2bs^2h$  & \multicolumn{1}{c|}{$6bsh$}    & $6bsh$  \\ \cline{3-8} 
            &                            & \multicolumn{2}{c|}{softmax}            & \multicolumn{1}{c|}{-}      & -      & \multicolumn{1}{c|}{$2bs^2$}    & $2bs^2$  \\ \cline{3-8} 
            &                            & \multicolumn{2}{c|}{attention on value} & \multicolumn{1}{c|}{$2bs^2h$}  & $2bs^2h$  & \multicolumn{1}{c|}{$2bs^2$}    & $2bs^2$  \\ \cline{3-8} 
            &                            & \multicolumn{2}{c|}{O Projection}             & \multicolumn{1}{c|}{$2bsh^2$}  & $2bsh^2$  & \multicolumn{1}{c|}{$2bsh$}    & $2bsh$  \\ \cline{2-8} 
            & norm                       & \multicolumn{2}{c|}{-}                  & \multicolumn{1}{c|}{$bsh$}    & $bsh$    & \multicolumn{1}{c|}{$2bsh$}    & $2bsh$  \\ \cline{2-8} 
            & \multirow{4}{*}{MLP}       & \multicolumn{1}{c|}{-}        & router  & \multicolumn{1}{c|}{-}      & -      & \multicolumn{1}{c|}{-}       & -     \\ \cline{3-8} 
            &                            & \multicolumn{1}{c|}{linear}   & linear  & \multicolumn{1}{c|}{$4bshg_d$} & $4bshg_d$ & \multicolumn{1}{c|}{$2bsh$}    & $2bsh$  \\ \cline{3-8} 
            &                            & \multicolumn{1}{c|}{swiglu}   & swiglu  & \multicolumn{1}{c|}{$bsg_d$}   & $bsg_e$   & \multicolumn{1}{c|}{$bsg_d$}    & $bsg_e$  \\ \cline{3-8} 
            &                            & \multicolumn{1}{c|}{linear}   & linear  & \multicolumn{1}{c|}{$2bshg_d$} & $2bshg_e$ & \multicolumn{1}{c|}{$bsg_d$}    & $bsg_e$  \\ \hline
Head        & -                          & \multicolumn{1}{c|}{-}        & -       & \multicolumn{1}{c|}{$2bshV$}  & $2bshV$  & \multicolumn{1}{c|}{$bsh$}     & $bsh$   \\ \hline
\end{tabular}}
\end{table}

%% file: autotuningsys.tex
\section{MoFa Tuning System}
In this section, we introduce the MoFa-based tuning system to identify various metrics corresponding to optimal training performance. We develop two tuning models targeting two distinct objectives: step-level optimization and E2E optimization. The step-level optimization model solely considers $\mathrm{CostModel}_{Mo}$, aiming to minimize the step time consumed per global batch. The E2E optimization model incorporates $\mathrm{CostModel}_{Fa}$ for joint optimization, targeting the minimization of the total time required to process the entire dataset.

\begin{figure}[t]
\centering
    \includegraphics[scale=0.8]{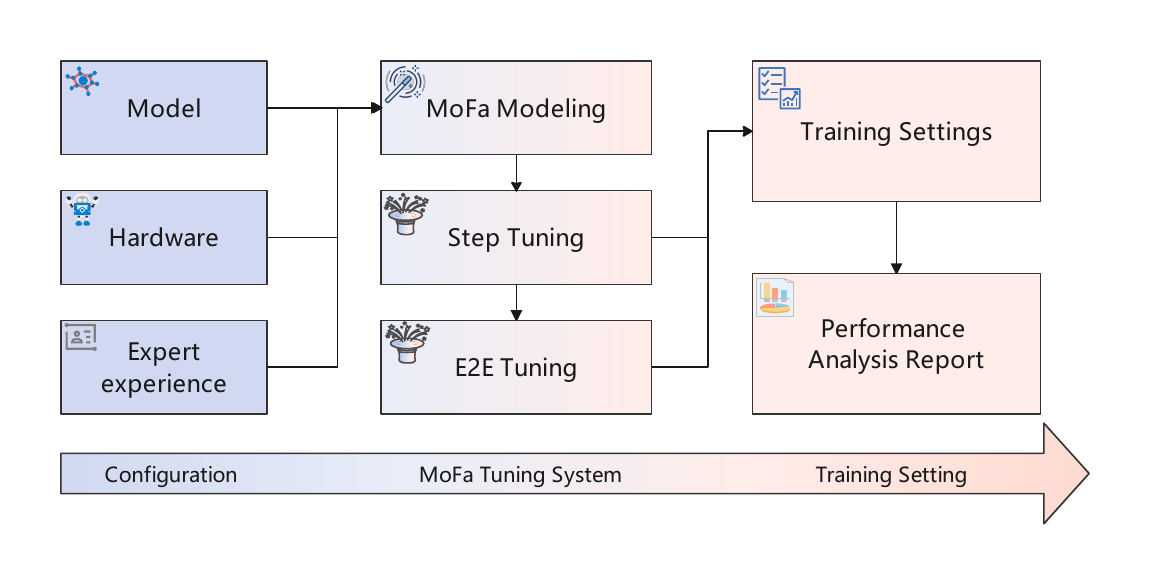}
    \caption{MoFa Tuning System.}
    \label{mofa-structure}
\end{figure}

\subsection{Step-level Tuning}

In distributed training of large models, the complete dataset is typically partitioned into multiple steps for training, with the input data for each step referred to as a global batch. We define the step-level optimization objective as minimizing the time consumption per step to achieve maximum step-level throughput. Consequently, the step-level optimization model is formulated as follows:
\begin{align}
\label{steptuning}
\nonumber
    & \min\limits_{t,c,p,e,d,m_{bs},LO,PO,OO,MO} \mathcal{G} = T^\mathrm{step} \\
    & s.t.\quad M^{\mathrm{Peak}} \leqslant M^\mathrm{GPU},\quad tcped \leqslant g_n,\quad m_{bs} \leqslant g_{bs}
\end{align}

Evidently, the solution space for \eqref{steptuning} is enormous. To accelerate the search and solution process, we propose a depth-first search (DFS) algorithm augmented with expert-knowledge-based pruning. The expert knowledge incorporates the following principles:
\begin{enumerate}
    \item Hardware resource constraints, as specified in \eqref{steptuning};
    \item Constraint among $g_{bs}$, $m_{bs}$, and $d$: $g_{bs}$ must be divisible by $(m_{bs} \times d)$;
    \item To reduce the PP bubble, the number of micro batches in the pipeline should satisfy $\frac{g_{bs}}{m_{bs}} \geqslant p$;
    \item TP communication is generally confined within a node, i.e., $t \leqslant N$, to minimize TP communication latency.
\end{enumerate}

By integrating these expert principles into the DFS algorithm, we can rapidly identify the parameter combination that delivers optimal step-level performance, thereby providing theoretical guidance for practical training.

\subsection{E2E-level Tuning}



During the model training process, the checkpoint saving interval also significantly impacts the E2E time consumption. Therefore, in E2E tuning, besides the parameters for step tuning, the model saving interval $I_{save}$ needs to be additionally determined. Consequently, the overall objective of E2E tuning is to minimize the total training time.

\begin{align}
\label{goal}
    \min_{t,c,p,e,d,m_{bs},LO,PO,OO,MO,I^\mathrm{ckpt}} \mathcal{G}&=\frac{ST^\mathrm{step}\left(1+\frac{T^\mathrm{save}}{I^\mathrm{ckpt}T^\mathrm{step}}\right)}{1-N_{nodes}r_f\left(u_b+\frac{I^\mathrm{ckpt}T^\mathrm{step}}{2}\right)}\\
     \nonumber s.t. M^{Peak}\leqslant & M^\mathrm{GPU},tcped\leqslant g_n, m_{bs} \leqslant g_{bs}
\end{align}

Note that in \eqref{goal}, when $T^\mathrm{step}$ decreases, $\mathcal{G}$ also decreases monotonically. Therefore, we can first solve for the optimal $T^\mathrm{step}$, and then individually optimize $I_{step}$. By taking the derivative of \eqref{goal}, the optimal saving interval can be found as:
\begin{align}
I^\mathrm{ckpt}=\frac{1}{T^\mathrm{step}}\left(-T^\mathrm{save}+\sqrt{{T^\mathrm{save}}^2-2T^\mathrm{save}\times u_b+\frac{2T^\mathrm{save}}{N_{nodes}r_f}}\right)
\end{align}

Then, round  $I^\mathrm{ckpt}$ both up and down, and select the one that minimizes $\mathcal{G}$.

\input{Table/5/GPUs}
\input{Table/5/modelchoice}




%% file: Table/5/GPUs.tex
\begin{table}[h!]
\caption{Hardware Specifications}
\renewcommand{\arraystretch}{1.3}
\label{hardwarechocie}
\centering
\scalebox{0.9}{
\begin{tabular}{l|l|l|l|l|l}
\hline
Device & $M^\mathrm{GPU}$ & $P^\mathrm{GPU}$ &$B^\mathrm{H2D}$ & $B^\mathrm{D2H}$ & $M^\mathrm{CPU}$\\ \hline
A & 32& 512& 32& 32 & 2 \\ \hline
B & 64 & 1024 & 32 & 32 & 2     \\ \hline
\end{tabular}}
\end{table}

%% file: Table/5/modelchoice.tex

\begin{table}[h!]
\centering
\caption{The configurations of the models in experiments.}
\label{modelchoice}
\scalebox{0.85}{
\begin{tabular}{cccc}
\hline
\diagbox{Config}{Model} & LLaMA2-70B & LLaMA3-405B & DeepSeek-V3 \\ \hline
$L$    & 80         & 126         & 61          \\
$s$    & 4096       & 8192        & 4096        \\
$h$    & 8192       & 16384       & 7168        \\
$a$    & 64         & 128         & 128         \\
$q$    & 8          & 16          & -          \\
$g_d$   & 28672      & 53248       & 18432       \\
$g_e$   & -         & -          & 2048        \\
$t_k$ & -         & -          & 8           \\
$V$    & 32000      & 128000      & 129280      \\
$r$             & -         & -          & 1536        \\
Attention  Type     & GQA        & GQA         & MLA         \\ \hline
\end{tabular}}
\end{table}

%% file: experiment.tex
\section{Experiment and Analysis}
\subsection{Experimental Setup}
Our experiments were conducted on two types of GPUs with detailed specifications listed in the Table~\ref{hardwarechocie}.
The distributed framework employed was Megatron-LM, and the software environment consisted of Python 3.8 and PyTorch 2.1.0.

\subsection{MoFa Performance Modeling Evaluation}
This section evaluates the modeling accuracy of MoFa. The evaluation process primarily selects the following three representative models: LLaMA2-70B, LLaMA3-405B, and DeepSeek-V3-671B. LLaMA2-70B and LLaMA3-405B are mainstream dense models below and above the 100B parameter scale, respectively, while DeepSeek-V3-671B is a prominent MoE model. Table~\ref{modelchoice} presents the key architectural parameters of these three models.

We utilized MoFa to model these three models and compared the results with actual performance metrics. As shown in Table~\ref{tab:pme}, the inputs to the MoFa modeling include the cluster hardware configuration, the adopted model parallel strategy, the optimization features enabled, and the fault tolerance configuration. The Performance Evaluation column displays the modeling results from MoFa. For LLaMA2-70B, performance was modeled across four cluster scales with 128, 256, 512, and 1024 GPUs. The results demonstrate a modeling accuracy of $99.6\%$ at the 128 GPUs scale and $97.65\%$ at the 1024 GPUs scale. The modeling accuracies for LLaMA3-405B and DeepSeek-V3-671B reach $98.73\%$ and $97.72\%$, respectively, indicating the high precision of MoFa.

Furthermore, we observed that the ETTR shows an increasing trend as the training scale expands. For instance, as the scale for LLaMA2-70B increases from 128 to 1024 GPUs, the ETTR improves from $99.49\%$ to $99.39\%$. Moreover, the rate of ETTR increase gradually slows with larger training scales. The ETTR increase is $0.6\%$ when scaling from 128 to 256 GPUs, but only $0.07\%$ when scaling from 512 to 1024 GPUs. This phenomenon occurs because, at the current training scales, the positive benefits from the increase in $t_{step}$ and the decrease in $t_{save}$ outweigh the negative impacts from the increased failure rate due to a larger $W_{n}$ and the increase in $u_b$. However, as the cluster size grows further, the proportion of negative impacts becomes increasingly significant, leading to a deceleration in the ETTR growth rate.
\input{Table/5/modeling-perf}

\subsection{Step-level Tuning Evaluation}
\input{Table/5/search-perf}
In practical training scenarios, practitioners typically train on clusters of fixed size. To maintain loss consistency, a fixed global batch size (GBS) is often used. This section uses MoFa tuning system to identify the optimal performance configuration, including the corresponding parallel strategy, optimization features, and key stage information, under constraints of fixed cluster size and GBS. Using the LLaMA3-405B model as an example, we selected three cluster scales for optimization: a hundred-GPU scale (64 GPUs), a thousand-GPU scale (2048 GPUs), and a ten-thousand-GPU scale (8192 GPUs). For each cluster scale, we considered two GBS settings: a small GBS and a large GBS, configured as 0.25 times and 1 times of the cluster size, respectively.

In the MoFa Tuning results, we present the top 4 optimal configurations for each scenario, as shown in Table~\ref{tuningresults}. Through MoFa Tuning, trainers can select the optimal configuration based on their specific context, thereby maximizing cluster resource utilization and accelerating the overall training process. For the small GBS scenario on the hundred-GPU scale cluster, e.g. 64 GPUs, modeling results indicates that the Top 1 configuration yields performance improvements of $6.24\%$, $13.22\%$, and $20.27\%$ compared to the configurations of Top 2, Top 3, and Top 4, respectively. Further analysis of key metrics reveals that the primary advantage of Top 1 over Top 2 stems from a larger $M_{bs}$. While increasing $M_{bs}$ augments PP latency, the reduction in computation operator latency yields a greater benefit. The advantage of Top 1 over Top 3 primarily comes from lower computation and TP latency, and its advantage over Top 4 arises from reduced computation and PP latency. 

Furthermore, optimization becomes more complex at the thousand-scale clusters even ten-thousand-scale clusters because of the DP. This analysis underscores the highly coupled nature of the training system, where hardware resources, model architecture, parallel strategies, and optimization features interact intricately, with each factor exerting varying influences on different stages, collectively determining the step-level performance. The MoFa tuning system performs step-level performance optimization aimed at minimizing training time and systematically identifies correlations among these elements and their impact on final system performance, thereby demonstrating the advanced capabilities and effectiveness of MoFa.

\subsection{E2E-level Tuning Evaluation}
\begin{figure}[t]
    \centering
    \includegraphics[scale=0.4]{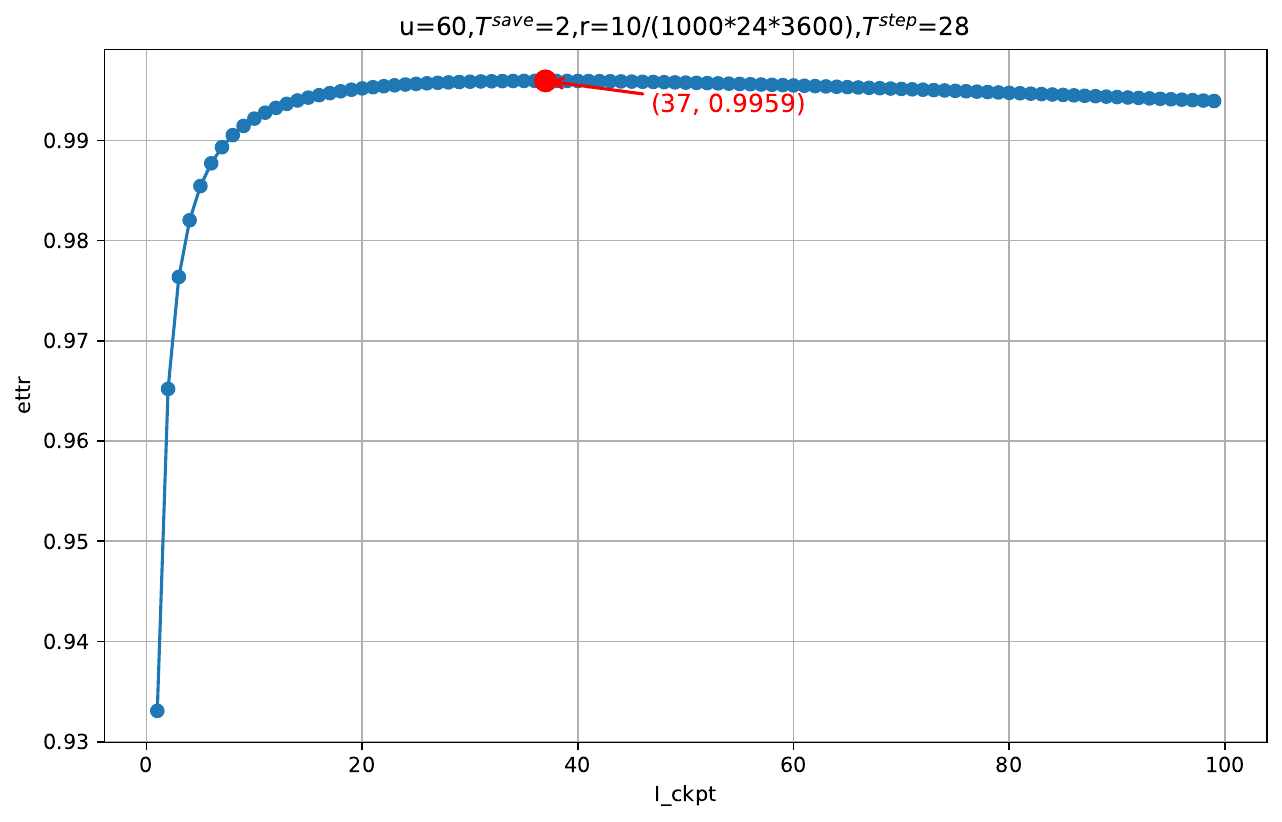}
    \caption{Trend of ETTR with $I_{ckpt}$.}
    \label{Ickpt&e2e}
\end{figure}

In this section, we primarily explore the impact of fault-tolerance-related parameters on E2E runtime. 


\subsubsection{Analysis of $I^\text{ckpt}$ on ETTR}

The $I^\text{ckpt}$ is a critical factor for E2E runtime. Unreasonable settings may result in increased time required for ckpt saving or increased training rollback time. Therefore, we further investigated the impact of $I^\text{ckpt}$ values on ETTR, as shown in Fig.\ref{Ickpt&e2e}. Assuming that the average single-fault recovery time $u_b=60$, the single ckpt saving time $T^\text{save}=2$, the single-node failure rate $r_f$=0.01 per node per day, and the number of nodes $N^\text{nodes} = 32$. ETTR increases with $I^\text{ckpt}$ at first and then decreases. In this scenario, the optimal $I^\text{ckpt}$ is 37, corresponding to an ETTR of 99.59$\%$.

\begin{figure}[t]
    \centering
    \subfigure[$N^\mathrm{nodes}$]{
    \includegraphics[width=0.48\linewidth]{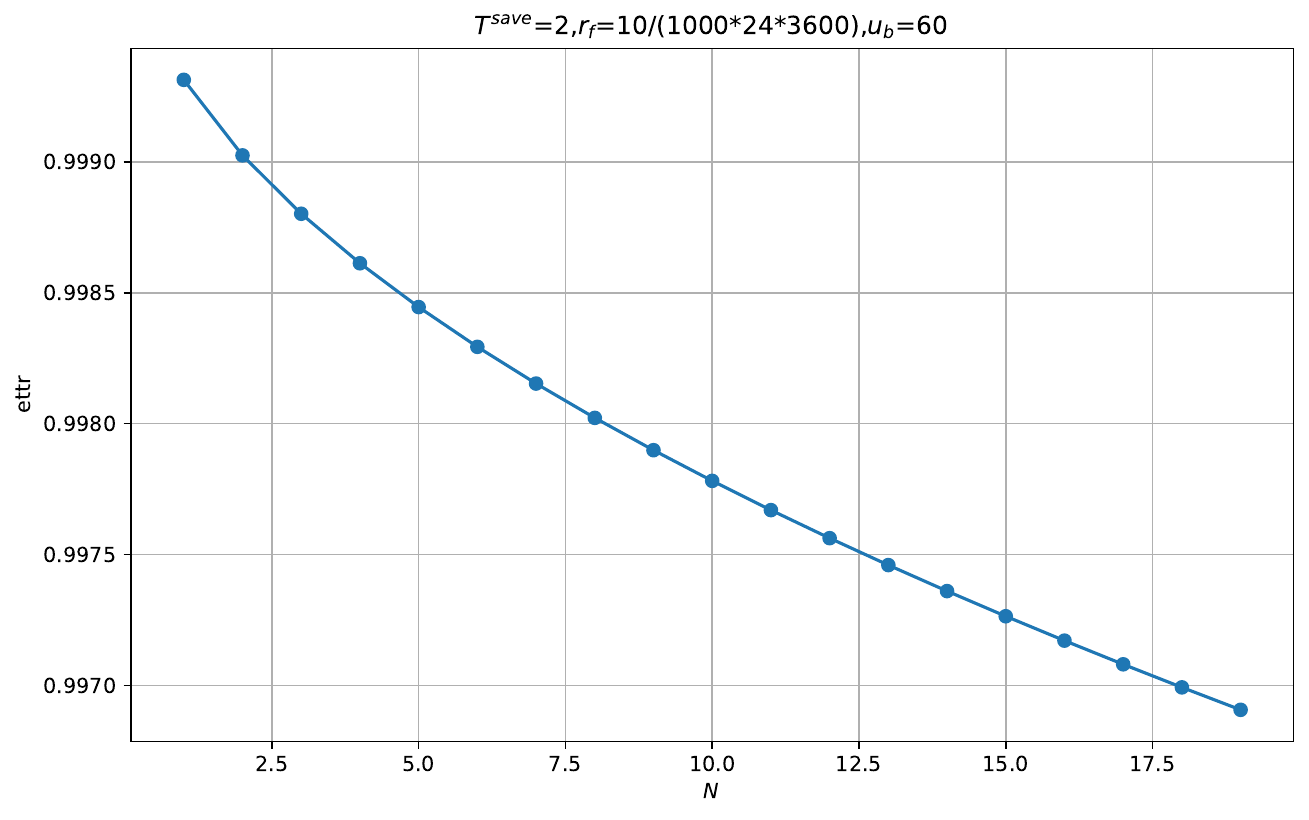}}
     \subfigure[$r_f$]{
     \includegraphics[width=0.48\linewidth]{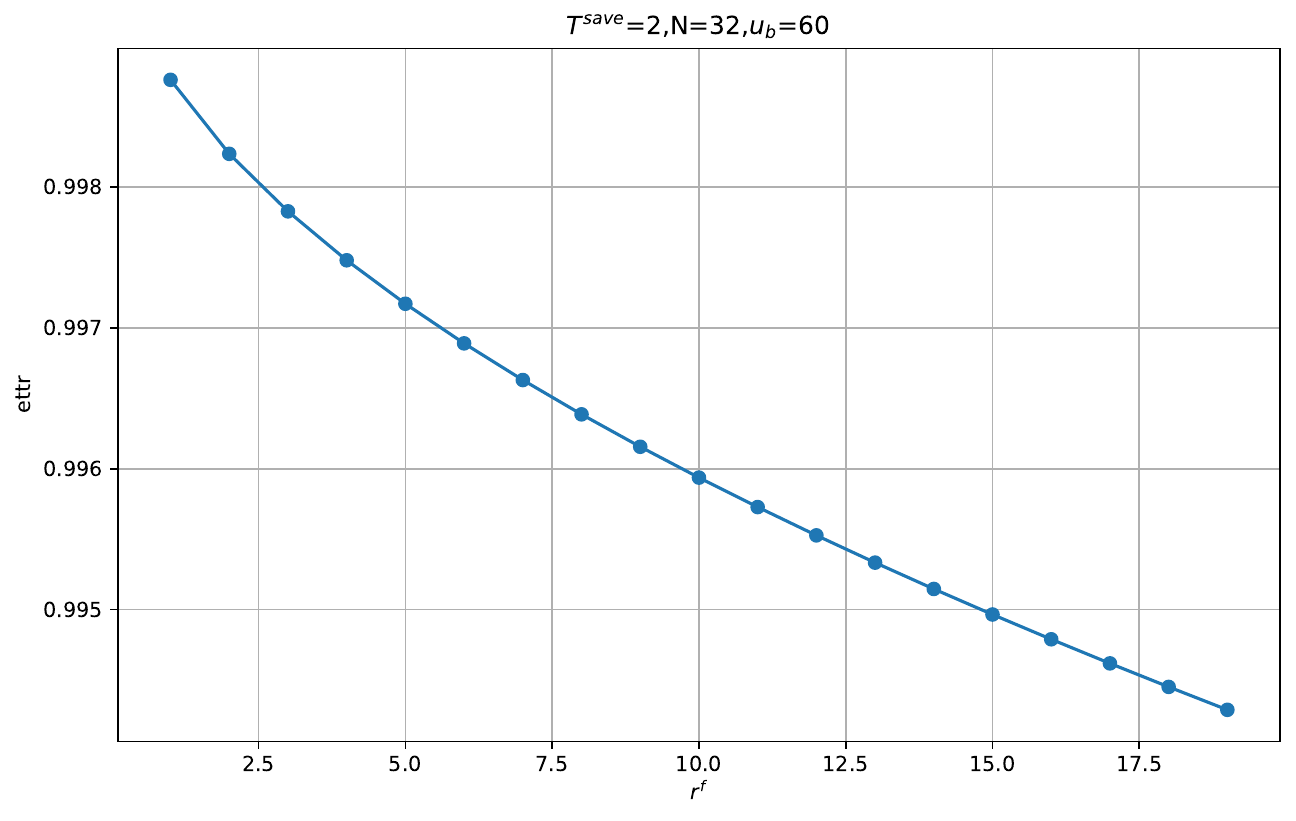}}

     \subfigure[$u_b$]{
     \includegraphics[width=0.48\linewidth]{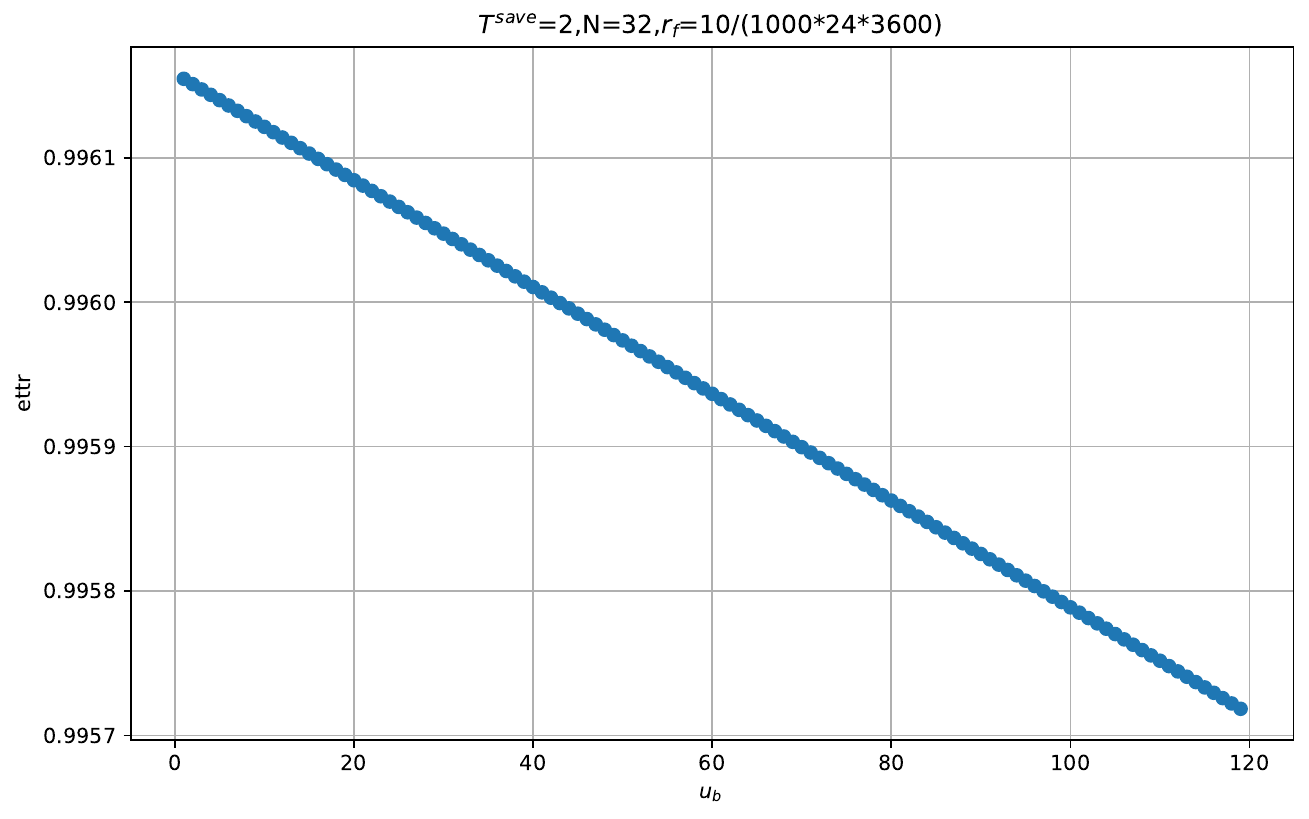}}
     \subfigure[$T^\mathrm{save}$]{
     \includegraphics[width=0.48\linewidth]{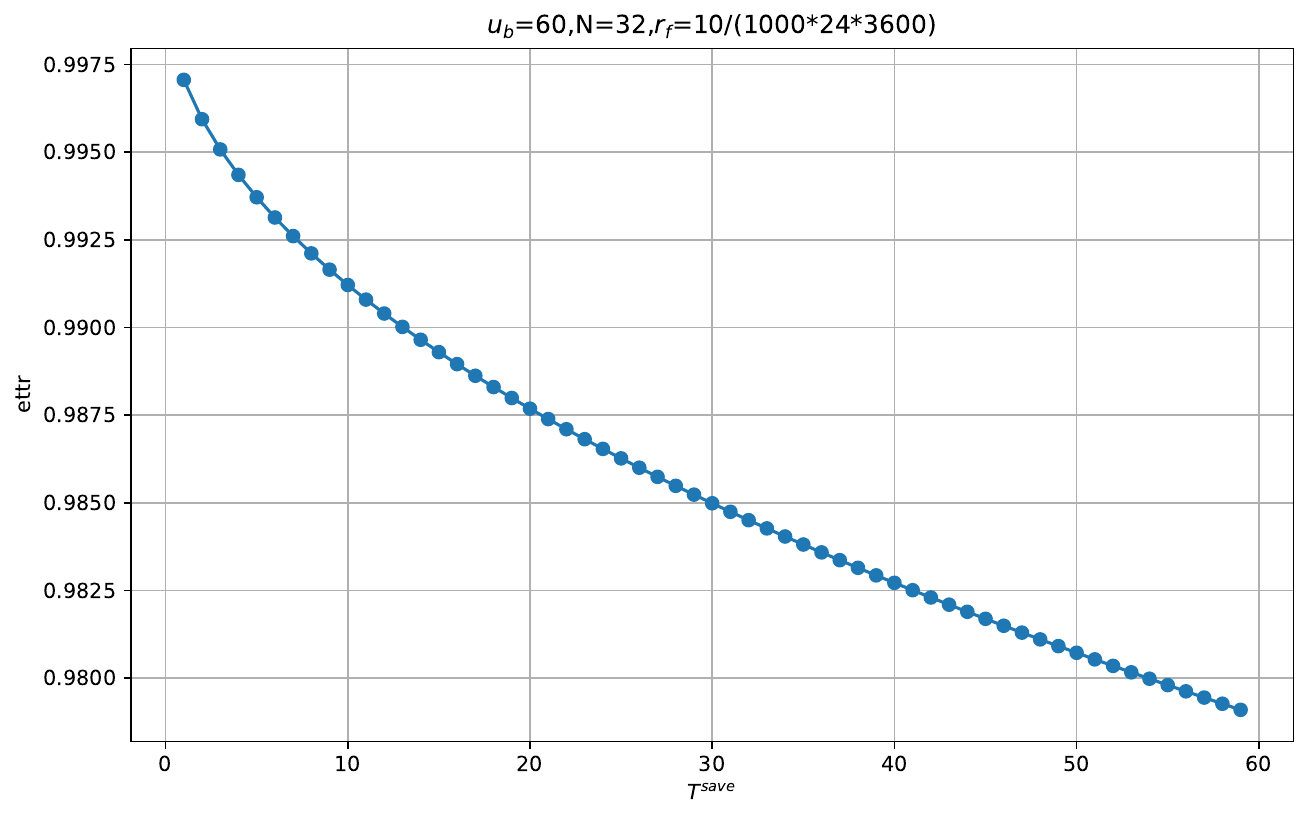}}
    \caption{
   Trend of ETTR with other factors.
    }
    \label{ettrfig}
    \end{figure}

\subsubsection{Analysis of other factors on ETTR}
In addition to the ckpt saving interval $I^\text{ckpt}$, we also investigate the impact of the single-node failure times $r_f$, the number of training nodes $N^\text{nodes}$, the single-fault average recovery time $u_b$, and the single ckpt saving time $T^\text{save}$ on ETTR. The results are shown in Fig.\ref{ettrfig}. It can be observed that holding other factors constant, when $N^\text{nodes}$ increases from 8 to 240, ETTR decreases from 99.80$\%$ to 98.80$\%$; when $r_f$ increases from 0.0025 to 0.0150, ETTR decreases from 99.80$\%$ to 99.50$\%$; When $u_b$ increases from 18 to 120, ETTR decreases from 99.61$\%$ to 99.572$\%$; when $T^\text{save}$ increases from 2 to 60, ETTR decreases from 99.50$\%$ to 97.50$\%$. This study provides valuable guidance for enhancing cluster utilization efficiency.

\subsection{Analysis of Key Settings on Performance}
This section investigates the impact of key configurations on performance during optimization, using the number of chunks $v$ and the optimizer type as examples.
\subsubsection{Effect of $v$}
\begin{figure}[t]
    \centering
    \includegraphics[scale=0.45]{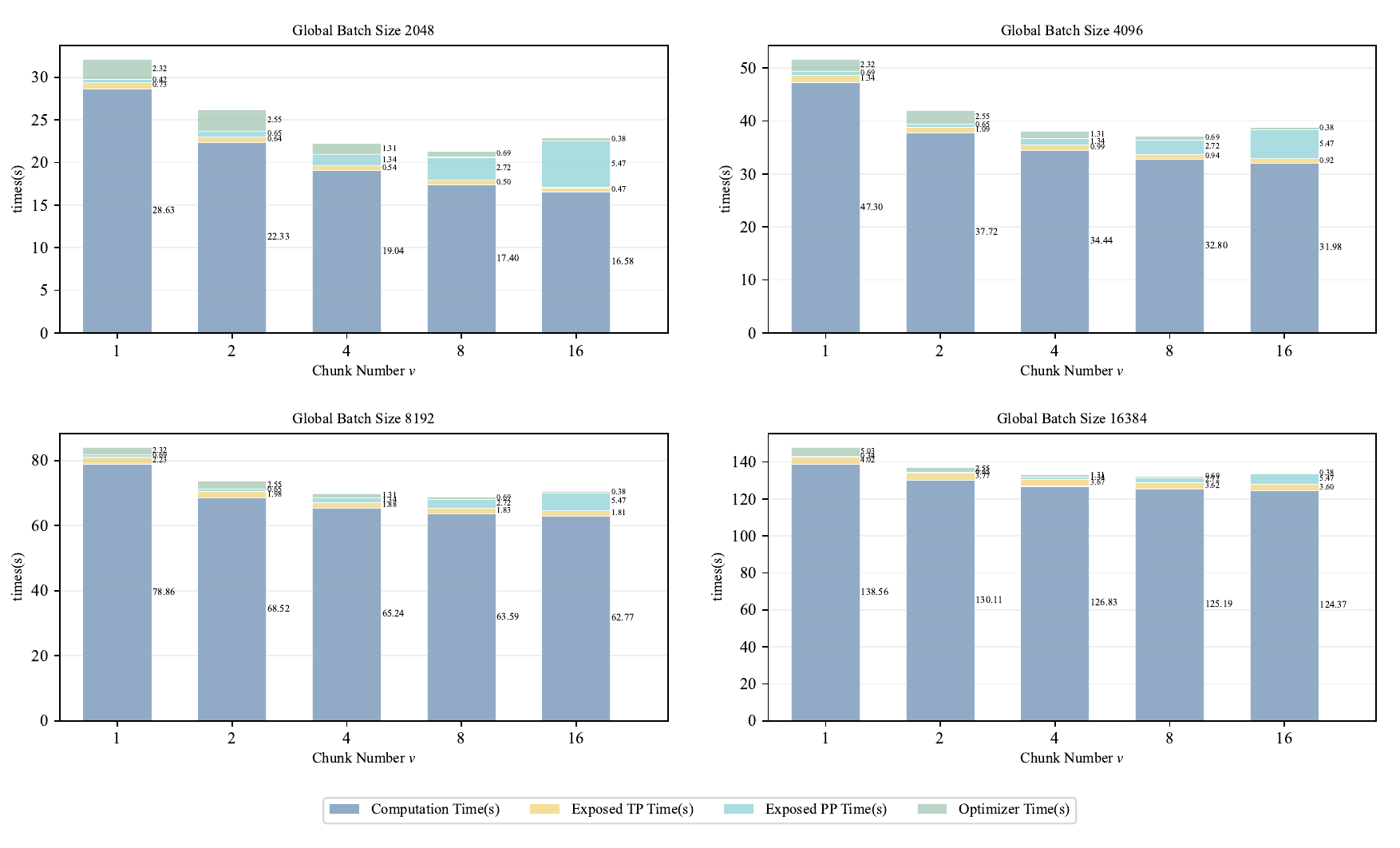}
    \caption{Effect of $v$.}
    \label{v-effect}
\end{figure}
Typically, to minimize the PP Bubble in the 1F1B schedule (the computational idle time due to pipeline stalls), practitioners set $v$ to its maximum value, i.e., $v=L/p$. However, according to \eqref{ppwarmup}–\eqref{ppcooldown}, increasing $v$ also multiplies the number and latency of PP communications by a factor of $v$. Although the steady phase can employ PP Overlap to mitigate PP communication latency, the PP overhead during the warmup and cooldown phases still increases. Consequently, increasing $v$ may not always be optimal. To investigate the effect of $v$, we used MoFa to model the training performance of LLaMA3-405B under different $v$ values, as shown in Fig.\ref{v-effect}. The GPU type was fixed to B, the number of GPUs to 2048, and four GBS scenarios were selected for analysis. Fig.\ref{v-effect} shows that the training latency under different GBS values initially decreases and then increases as $v$ grows, indicating that the optimal $v$ is 8 rather than 16. This is because the additional exposed PP latency at $v=16$ outweighs the reduction in PP Bubble. For example, with $G_{bs}=2048$, $v=16$ increases PP latency by 2.7 seconds compared to $v=8$, but only reduces the bubble by 1 second. Furthermore, as $G_{bs}$ increases, the latency difference between different $v$ values diminishes. This occurs because a larger $G_{bs}$ reduces the PP Bubble and decreases the proportional impact of the added PP latency from the warmup and cooldown phases.

\subsubsection{Effect of Optimizer Optimization}
Optimizer states can account for up to $80\%$ of static memory, necessitating memory optimizations. Different optimizer memory optimization strategies impact performance differently. This section examines the performance effects of the two optimizer memory optimizations described in \eqref{cpuadam-mem} and \eqref{distributed-optimizer-memory}. Using LLaMA3-405B training as an example, we selected two scenarios: single DP group at the hundred-GPU scale and multiple DP groups at the thousand-GPU scale, to provide a comprehensive analysis. Table \ref{tab:optimizer-effect} presents the parallel strategies for the Top 3 performance configurations for LLaMA3-405B with a single DP group at the hundred-GPU scale, using GPU type B and $G_{bs}=16$. It shows that with the CPU optimizer, LLaMA3-405B can be deployed on just 64 GPUs; otherwise, a minimum of 256 GPUs is required, which would introduce significant PP Bubble, degrading single-DP performance by approximately $50\%$. Concurrently, the CPU optimizer demonstrates higher memory optimization efficiency than the distributed optimizer, further reducing the memory required for a 64-GPU deployment compared to the 256-GPU case. Therefore, from the perspectives of single-DP performance and resource utilization, the CPU optimizer is a favorable choice.

\begin{figure}[t]
    \centering
    \includegraphics[scale=0.5]{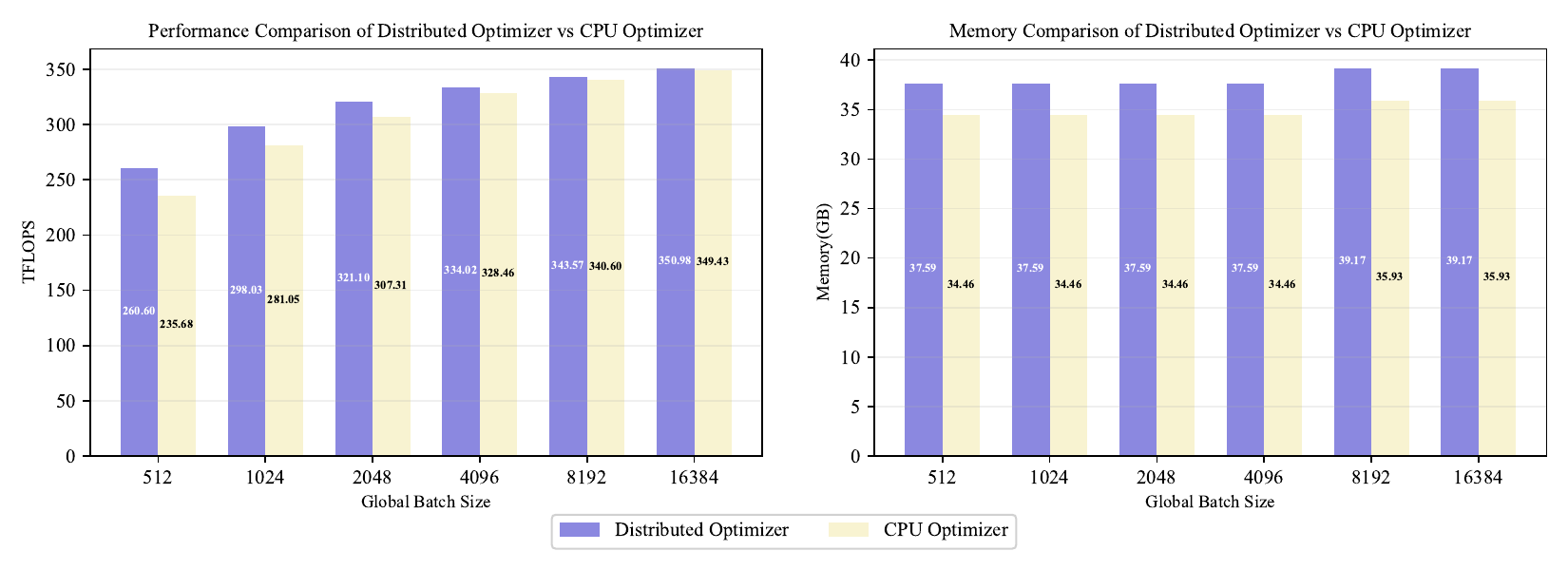}
    \caption{Distributed Optimizer vs CPU Optimizer.}
    \label{optimizer-compare}
\end{figure}

At the thousand-GPU scale, we analyzed the optimal performance and memory consumption between the CPU optimizer and the distributed optimizer under different GBS values, as shown in Fig.\ref{optimizer-compare}. The results indicate that the performance gap between the CPU optimizer and the distributed optimizer narrows as $G_{bs}$ increases, decreasing from $9.6\%$ to $0.4\%$. This is primarily because a larger $G_{bs}$ reduces the proportion of optimizer overhead. However, the right panel of Fig.\ref{optimizer-compare} shows that the CPU optimizer maintains a memory advantage over the distributed optimizer, saving approximately $8.3\%$ additional memory across all GBS values. Consequently, for large-scale training scenarios at the thousand or ten-thousand GPU scale with large GBS, employing the CPU optimizer offers better cost-effectiveness.

\input{Table/5/optimizereffect}

\subsection{Cluster Configuration on Performance}
As training cluster scales increase, cluster scalability and node architecture become critical to training performance. This section explores the impact of large-scale cluster linearity and node configuration on training performance based on MoFa.
\subsubsection{Cluster Size}
Using LLaMA3-405B as an example, this section investigates how cluster linearity changes with increasing scale and the impact of the DP Overlap feature on linearity through MoFa Tuning. Specifically, we used MoFa to find the optimal performance for clusters ranging from 64 to 8192 GPUs under two scenarios: $G_{bs}=1024$ and $G_{bs}=4096$. The cluster linearity was calculated using the following formula:
\begin{align}
    linearity = \frac{T_1^\mathrm{step}}{T_2^\mathrm{step}}
\end{align}
where $T_1^\mathrm{step}$ and $T_2^\mathrm{step}$ are the consumed time of per training step of cluster 1 and cluster 2, respectively.

\begin{figure}[t]
    \centering
    \begin{minipage}{0.48\textwidth}
        \centering
        \includegraphics[scale=0.38]{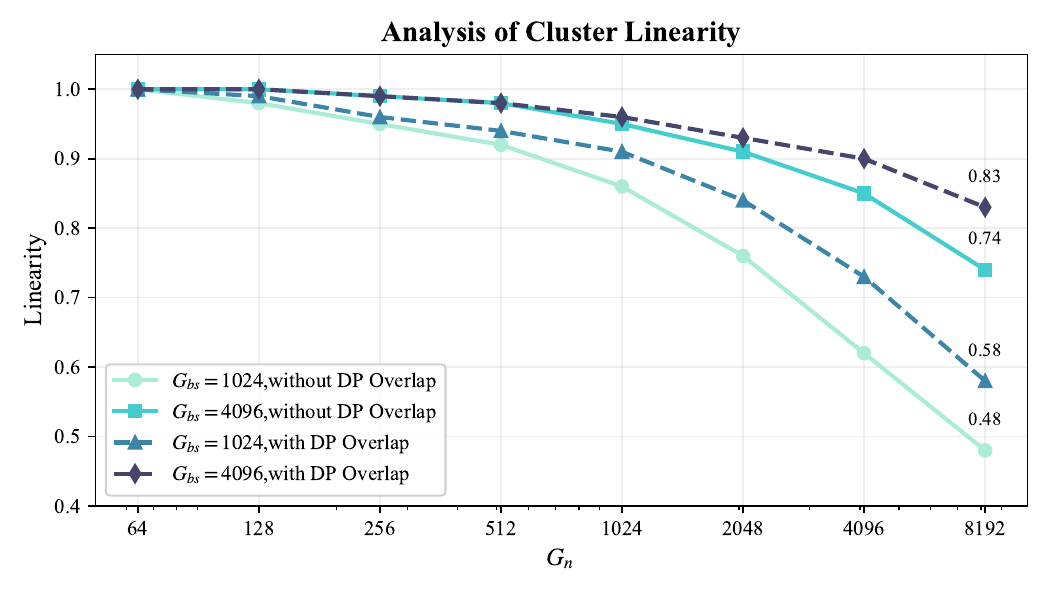}
        \caption{Analysis of Cluster Linearity.}
        \label{linearity-analysis}
    \end{minipage}
    \hfill
    \begin{minipage}{0.48\textwidth}
        \centering
        \includegraphics[scale=0.38]{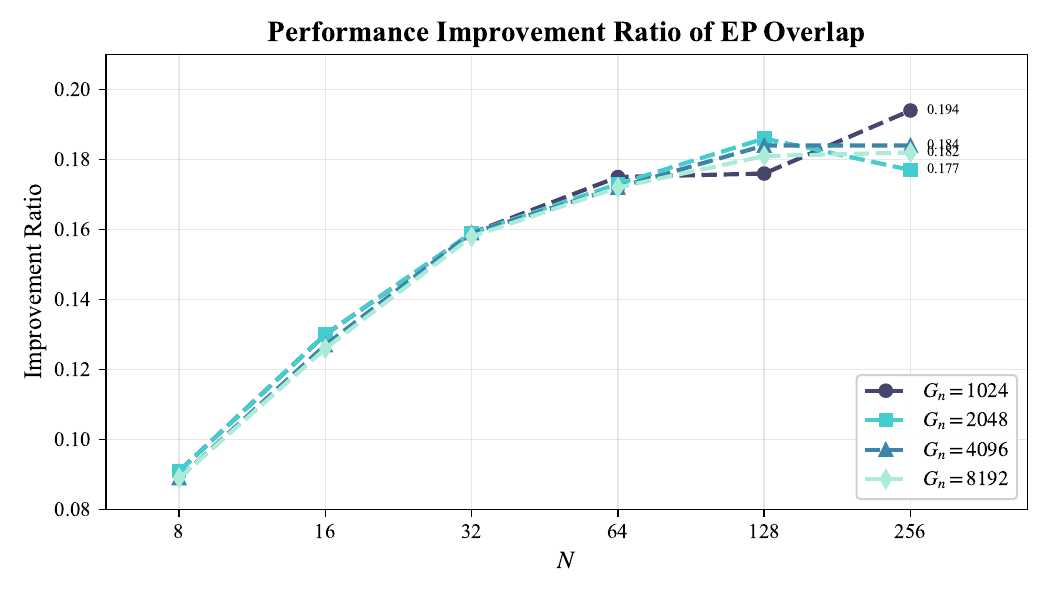}
        \caption{Performance Improvement Ratio of EP Overlap}
        \label{ep-overlap}
    \end{minipage}
\end{figure}

As illustrated in Fig.\ref{linearity-analysis}. The results show that increasing the $G_{bs}$ and enabling the DP Overlap feature effectively enhance cluster linearity. The linearity for $G_{bs}=4096$ with DP Overlap is improved by $9\%$, $25\%$, and $35\%$ compared to $G_{bs}=4096$ without DP Overlap, $G_{bs}=1024$ with DP Overlap, and $G_{bs}=1024$ without DP Overlap, respectively. In addition, we can see that even the cluster linearity is only $83\%$ under setting great GBS and applying DP overlap. The main reason is that DP communication cannot be completely overlap by computation, indicating that designing more effective overlap methods is a promising approach.


\input{Table/5/N-effect}

\subsubsection{Node Size}
As a key technology for reducing distributed communication, the design 
of supernode scale is promising research. This subsection uses the DeepSeek-V3-671B model and MoFa tuning system to illustrate the impact of supernode scale on the optimal EP settings and the effectiveness of the EP Overlap for MoE models.
On the one hand, Table \ref{tab:N-effect} presents the performance of optimal EP configurations tuned by MoFa under different supernode configurations for $G_n=1024, 2048, 4096, 8192$, with the EP Overlap optimization fixed as enabled. The results indicate that both the cluster scale and the supernode configuration jointly influence the optimal EP setting. Specifically, for $G_n=1024$ and $2048$, the optimal EP is $e=128$, while for $G_n=4096$ and $8192$, the optimal EP is $e=256$. A larger EP generally leads to higher expert computation efficiency, as evidenced by the $T^\mathrm{cal}$ metric. Furthermore, the $T^\mathrm{step}$ values are closed when the supernode size ranges from $N=32$ to $256$, with differences less than $4\%$. This suggests the presence of a marginal effect in supernode scaling, where the efficiency gains diminish as the supernode size increases.
On the other hand, Fig.\ref{ep-overlap} explores the performance improvement from EP Overlap under different $N$ values. The results show that the benefit of EP Overlap increases with larger $N$. This is primarily because increasing $N$ can reduce expert computation latency, thereby increasing the proportion of EP communication overhead within a training step. Then, applying EP overlap can reduce this EP communication overhead, yielding greater benefits. Moreover, this conclusion holds across the evaluated range of $G_n=1024$ to $8192$, with performance gains exceeding $17\%$ particularly for $N$ between 64 and 256.

\subsection{Efficiency of Computational Efficiency}
\begin{figure}[t]
    \centering
    \includegraphics[scale=0.5]{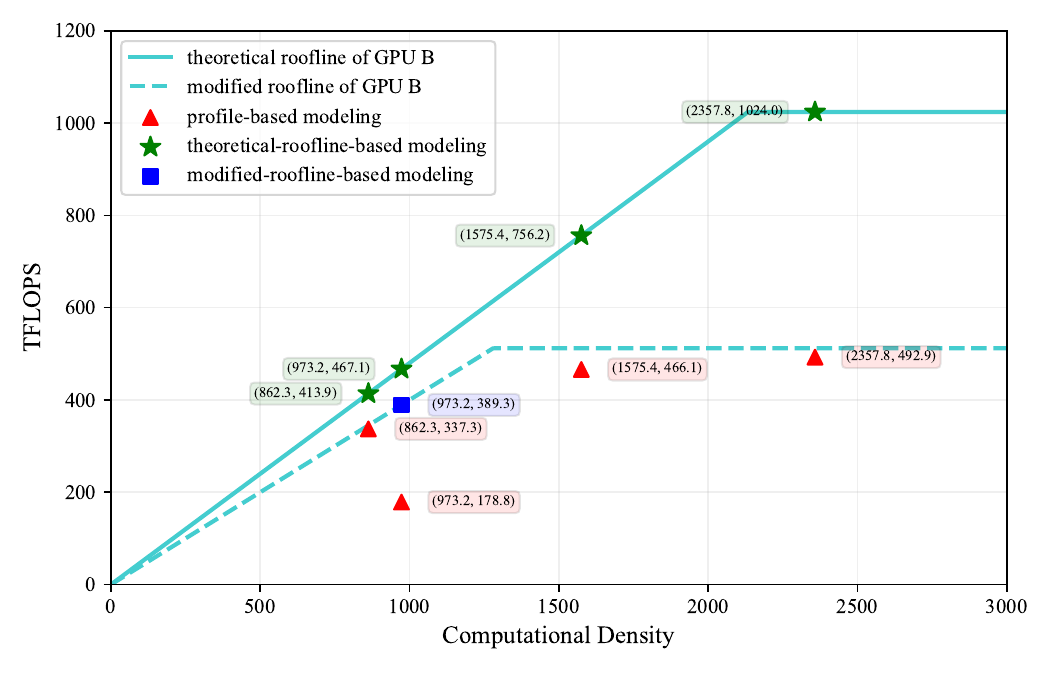}
    \caption{Analysis of roofline.}
    \label{roofline}
\end{figure}

As the cornerstone of training performance, communication efficiency is crucial to overall training performance. This section investigates the impact of three approaches of communication modeling on training performance: profile-based modeling, theoretical-roofline modeling, and modified-roofline modeling. Take GPU B as an example, its modified-roofline model is obtained from linear fitting with about 1,000 data points. As shown in Fig.\ref{roofline}, we plot the theoretical roofline (solid green line) and the modified roofline (dashed green line) for GPU B. We also marked the profile-based datas of main matrix multiply (MM) for LLaMA3-405B, including the Q-K-V fused MM, attention, O projection, and MLP, corresponding to the red triangles.

It can be observed that the efficiencies of Q-K-V fused MM, O projection, and MLP are close to the modified-roofline performance, indicating that their current implementations are near the hardware’s practical limit. In contrast, the attention operator shows a significant performance gap compared to the modified roofline, suggesting implementation flaws that need to be addressed for optimization. Furthermore, the attention, O projection, and MLP all exhibit considerable gaps relative to the theoretical roofline, indicating that further algorithmic innovations are required to improve communication efficiency. This provides a methodology for identifying optimizable operators.

On the other hand, we identify the point (937.2, 178.8) in Fig.\ref{roofline} as an outlier. Assuming that this outlier can be optimized to (937.2, 389.3), we analysis its impact for MoFa tuning, as summarized in Table~\ref{tab:profile-roofline}, where the settings of cluster scale and GBS are consistent with those in Table~\ref{steptuning}. The results show that the optimal training setting remains same after optimizing the outlier. Specially, the $117.73\%$ improvement in outlier efficiency benifit a $2\% \sim 3\%$ step performance gain under this optimal setting.

\input{Table/5/roofline-profile}

%% file: Table/5/modeling-perf.tex
\begin{table}[t]
\centering
\caption{MoFa Modeling Evaluation}
\label{tab:pme}
\renewcommand{\arraystretch}{1.3}
\scalebox{0.55}{
\begin{tabular}{cccccccc}
\hline
\multirow{6}{*}{Configuration} & Model & LLaMA2-70B & LLaMA2-70B & LLaMA2-70B & LLaMA2-70B & LLaMA3-405B & DeepSeek-V3-671B \\
 & GPU & A & A & A & A & B & B \\
 & $M^\mathrm{GPU}$ & 32 & 32 & 32 & 32 & 64 & 64 \\
 & $G_n$ & 128 & 256 & 512 & 1024 & 64 & 128 \\
 & $N$ & 8 & 8 & 8 & 8 & 8 & 32 \\
 & $D_s$(T) & 15 & 15 & 15 & 15 & 15 & 15 \\ \hline
\multirow{7}{*}{Parallelism} & $t$ & 8 & 8 & 8 & 8 & 8 & 1 \\
 & $c$ & 1 & 1 & 1 & 1 & 1 & 1 \\
 & $e$ & 1 & 1 & 1 & 1 & 1 & 32 \\
 & $p$ & 8 & 8 & 8 & 8 & 8 & 4 \\
 & $d$ & 2 & 4 & 8 & 16 & 1 & 1 \\
 & $G_{bs}$ & 256 & 512 & 1024 & 2048 & 16 & 960 \\
 & $M_{bs}$ & 2 & 2 & 2 & 2 & 1 & 1 \\ \hline
\multirow{4}{*}{Optimization} & $LO$ & \eqref{suanzi-up}, \eqref{tongxin-up} & \eqref{suanzi-up}, \eqref{tongxin-up} & \eqref{suanzi-up}, \eqref{tongxin-up} & \eqref{suanzi-up}, \eqref{tongxin-up} & \eqref{suanzi-up}$\sim$\eqref{tp-overlap-costmodel} & \eqref{suanzi-up},\eqref{tongxin-up},\eqref{ep-overlap-costmodel} \\
 & $PO$ & \eqref{PPOverlap} & \eqref{PPOverlap} & \eqref{PPOverlap} & \eqref{PPOverlap} & \eqref{PPOverlap} & \eqref{PPOverlap} \\
 & $OO$ & - & - & - & - & - & - \\
 & $MO$ & \eqref{distributed-optimizer-memory},\eqref{distributed-optimizer-performance},\eqref{offload-mem}$\sim$\eqref{offload-bwd} & \eqref{distributed-optimizer-memory},\eqref{distributed-optimizer-performance},\eqref{offload-mem}$\sim$\eqref{offload-bwd} & \eqref{distributed-optimizer-memory},\eqref{distributed-optimizer-performance},\eqref{offload-mem}$\sim$\eqref{offload-bwd} & \eqref{distributed-optimizer-memory},\eqref{distributed-optimizer-performance},\eqref{offload-mem}$\sim$\eqref{offload-bwd} & \eqref{cpuadam-mem},\eqref{cpuadam-performance},\eqref{offload-mem}$\sim$\eqref{offload-bwd} & \eqref{cpuadam-mem},\eqref{cpuadam-performance},\eqref{full-mem},\eqref{full-performance} \\ \hline
\multirow{4}{*}{Tolerance} & $r_f$ & 0.50\% & 0.50\% & 0.50\% & 0.50\% & 0.50\% & 0.50\% \\
 & $T^\mathrm{save}$(s) & 4.19 & 2.35 & 1.59 & 0.95 & 9.3 & 7.7 \\
 & $I^\mathrm{ckpt}$ & 10 & 10 & 10 & 10 & 10 & 10 \\
 & $u_b$ & 134.41 & 147.72 & 174.34 & 227.58 & 127.75 & 134.41 \\ \hline
\multirow{9}{*}{Performance} & $T^\mathrm{step}$(s) & 27.83 & 27.99 & 28.33 & 28.83 & 24.46 & 74.5 \\
 & $\mathrm{TFLOPS}$ & 123.6 & 122.9 & 121.4 & 119.3 & 213.01 & 105.4 \\
 & $T^\mathrm{step}_\mathrm{active}$(s) & 27.94 & 28.1 & 28.54 & 29.52 & 24.77 & 76.24 \\
 & $\mathrm{TFLOPS}_\mathrm{active}$ & 123.1 & 122.4 & 120.5 & 116.5 & 210.3 & 103.1 \\
 & \textbf{Accuracy} & \textbf{99.60\%} & \textbf{99.59\%} & \textbf{99.26\%} & \textbf{97.65\%} & \textbf{98.73\%} & \textbf{97.72\%} \\
 & $S$ & 953675 & 476838 & 238419 & 119210 & 15258790 & 254314 \\
 & \textbf{ETTR} & 98.49\% & 99.11\% & 99.32\% & 99.39\% & 96.32\% & 98.96\% \\
 & $T^\mathrm{E2E}$(s) & 26947190.75 & 13465926.21 & 6800277.57 & 3457670.27 & 387465533.9 & 19144461.2 \\ \hline
\end{tabular}}
\end{table}



%% file: Table/5/search-perf.tex
\begin{table}[t]
\centering
\caption{Results of MoFa Tuning System.}
\label{tuningresults}
\renewcommand{\arraystretch}{1.3}
\scalebox{0.55}{
\begin{tabular}{cc|ccccc|cccc|ccc|ccccc}
\hline
\multicolumn{2}{c}{\textbf{Input}} & \multicolumn{5}{c}{\textbf{Parallelism}} & \multicolumn{4}{c}{\textbf{Optimization}} & \multicolumn{3}{c}{\textbf{Performance}} & \multicolumn{5}{c}{\textbf{Key Stage}} \\
$G_n$ & $G_bs$ & $t$ & $c$ & $p$ & $d$ & $M_{bs}$ & $LO$ & $PO$ & $OO$ & $MO$ & Memory & TFLOPS & $T^\mathrm{step}$ & $T^\mathrm{cal}$ & $T^\mathrm{TP}$ & $T^\mathrm{PP}$ & $T^\mathrm{DP}$ & $T^\mathrm{update}$ \\
\hline
\multirow{8}{*}{64} & 16 & 8 & 1 & 8 & 1 & 2 & \eqref{suanzi-up}$\sim$\eqref{tp-overlap-costmodel} & \eqref{PPOverlap} & - & \eqref{cpuadam-mem},\eqref{cpuadam-performance},\eqref{offload-mem}$\sim$\eqref{offload-bwd} & 34.46 & 241.18 & 22.86 & 17.40 & 0.50 & 2.72 & 0.00 & 2.25 \\
 & 16 & 8 & 1 & 8 & 1 & 1 & \eqref{suanzi-up}$\sim$\eqref{tp-overlap-costmodel} & \eqref{PPOverlap} & - & \eqref{cpuadam-mem},\eqref{cpuadam-performance},\eqref{offload-mem}$\sim$\eqref{offload-bwd} & 33.96 & 227.01 & 24.29 & 19.88 & 0.52 & 1.64 & 0.00 & 2.25 \\
 & 16 & 8 & 1 & 8 & 1 & 1 & \eqref{suanzi-up}, \eqref{tongxin-up}& \eqref{PPOverlap} & - & \eqref{cpuadam-mem},\eqref{cpuadam-performance},\eqref{offload-mem}$\sim$\eqref{offload-bwd} & 33.96 & 213.01 & 25.89 & 19.93 & 2.07 & 1.64 & 0.00 & 2.25 \\
 & 16 & 4 & 1 & 16 & 1 & 1 & \eqref{suanzi-up}$\sim$\eqref{tp-overlap-costmodel} & \eqref{PPOverlap} & - & \eqref{cpuadam-mem},\eqref{cpuadam-performance},\eqref{offload-mem}$\sim$\eqref{offload-bwd} & 35.43 & 200.54 & 27.50 & 19.54 & 0.23 & 5.47 & 0.00 & 2.25 \\
 & 64 & 8 & 1 & 8 & 1 & 2 & \eqref{suanzi-up}$\sim$\eqref{tp-overlap-costmodel} & \eqref{PPOverlap} & - & \eqref{cpuadam-mem},\eqref{cpuadam-performance},\eqref{offload-mem}$\sim$\eqref{offload-bwd} & 34.46 & 313.33 & 70.40 & 63.59 & 1.83 & 2.72 & 0.00 & 2.25 \\
 & 64 & 4 & 1 & 16 & 1 & 2 & \eqref{suanzi-up}$\sim$\eqref{tp-overlap-costmodel} & \eqref{PPOverlap} & - & \eqref{cpuadam-mem},\eqref{cpuadam-performance},\eqref{offload-mem}$\sim$\eqref{offload-bwd} & 35.93 & 297.92 & 74.04 & 65.18 & 0.83 & 5.77 & 0.00 & 2.25 \\
 & 64 & 8 & 1 & 8 & 1 & 2 & \eqref{suanzi-up}, \eqref{tongxin-up}& \eqref{PPOverlap} & - & \eqref{cpuadam-mem},\eqref{cpuadam-performance},\eqref{offload-mem}$\sim$\eqref{offload-bwd} & 34.46 & 289.97 & 76.07 & 63.76 & 7.34 & 2.72 & 0.00 & 2.25 \\
 & 64 & 4 & 1 & 16 & 1 & 2 & \eqref{suanzi-up}, \eqref{tongxin-up}& \eqref{PPOverlap} & - & \eqref{cpuadam-mem},\eqref{cpuadam-performance},\eqref{offload-mem}$\sim$\eqref{offload-bwd} & 35.93 & 287.94 & 76.61 & 65.26 & 3.32 & 5.77 & 0.00 & 2.25 \\
\hline
\multirow{8}{*}{2048} & 512 & 8 & 1 & 8 & 32 & 2 & \eqref{suanzi-up}$\sim$\eqref{tp-overlap-costmodel} & \eqref{PPOverlap} & \eqref{dpoverlap} & \eqref{cpuadam-mem},\eqref{cpuadam-performance},\eqref{offload-mem}$\sim$\eqref{offload-bwd} & 37.59 & 260.60 & 21.16 & 17.40 & 0.50 & 2.72 & 0.53 & 0.02 \\
 & 512 & 8 & 1 & 8 & 32 & 2 & \eqref{suanzi-up}, \eqref{tongxin-up}& \eqref{PPOverlap} & \eqref{dpoverlap} & \eqref{cpuadam-mem},\eqref{cpuadam-performance},\eqref{offload-mem}$\sim$\eqref{offload-bwd} & 37.59 & 243.02 & 22.69 & 17.44 & 1.98 & 2.72 & 0.53 & 0.02 \\
 & 512 & 8 & 1 & 8 & 32 & 2 & \eqref{suanzi-up}$\sim$\eqref{tp-overlap-costmodel} & \eqref{PPOverlap} & \eqref{dpoverlap} & \eqref{distributed-optimizer-memory},\eqref{distributed-optimizer-performance},\eqref{offload-mem}$\sim$\eqref{offload-bwd} & 34.46 & 235.68 & 23.40 & 17.40 & 0.50 & 2.72 & 0.53 & 2.25 \\
 & 512 & 8 & 1 & 8 & 32 & 1 & \eqref{suanzi-up}$\sim$\eqref{tp-overlap-costmodel} & \eqref{PPOverlap} & - & \eqref{cpuadam-mem},\eqref{cpuadam-performance},\eqref{offload-mem}$\sim$\eqref{offload-bwd} & 37.09 & 231.65 & 23.81 & 19.88 & 0.52 & 1.64 & 1.75 & 0.02 \\
 & 2048 & 8 & 1 & 8 & 32 & 2 & \eqref{suanzi-up}$\sim$\eqref{tp-overlap-costmodel} & \eqref{PPOverlap} & \eqref{dpoverlap} & \eqref{distributed-optimizer-memory},\eqref{distributed-optimizer-performance},\eqref{offload-mem}$\sim$\eqref{offload-bwd} & 37.59 & 321.10 & 68.70 & 63.59 & 1.83 & 2.72 & 0.53 & 0.02 \\
 & 2048 & 8 & 1 & 8 & 32 & 2 & \eqref{suanzi-up}$\sim$\eqref{tp-overlap-costmodel} & \eqref{PPOverlap} & \eqref{dpoverlap} & \eqref{cpuadam-mem},\eqref{cpuadam-performance},\eqref{offload-mem}$\sim$\eqref{offload-bwd} & 34.46 & 310.97 & 70.93 & 63.59 & 1.83 & 2.72 & 0.53 & 2.25 \\
 & 2048 & 4 & 1 & 16 & 32 & 2 & \eqref{suanzi-up}$\sim$\eqref{tp-overlap-costmodel} & \eqref{PPOverlap} & \eqref{dpoverlap} & \eqref{distributed-optimizer-memory},\eqref{distributed-optimizer-performance},\eqref{offload-mem}$\sim$\eqref{offload-bwd} & 39.17 & 304.93 & 72.34 & 65.18 & 0.83 & 5.77 & 0.53 & 0.02 \\
 & 2048 & 8 & 1 & 16 & 16 & 2 & \eqref{suanzi-up}$\sim$\eqref{tp-overlap-costmodel} & \eqref{PPOverlap} & \eqref{dpoverlap} & \eqref{distributed-optimizer-memory},\eqref{distributed-optimizer-performance},\eqref{offload-mem}$\sim$\eqref{offload-bwd} & 25.21 & 301.45 & 73.17 & 65.61 & 1.84 & 5.47 & 0.24 & 0.01 \\
\hline
\multirow{8}{*}{8192} & 2048 & 8 & 1 & 8 & 128 & 2 & \eqref{suanzi-up}$\sim$\eqref{tp-overlap-costmodel} & \eqref{PPOverlap} & \eqref{dpoverlap} & \eqref{cpuadam-mem},\eqref{cpuadam-performance},\eqref{offload-mem}$\sim$\eqref{offload-bwd} & 35.25 & 258.73 & 21.31 & 17.40 & 0.50 & 2.72 & 0.69 & 0.02 \\
 & 2048 & 8 & 1 & 8 & 128 & 2 & \eqref{suanzi-up}, \eqref{tongxin-up}& \eqref{PPOverlap} & \eqref{dpoverlap} & \eqref{cpuadam-mem},\eqref{cpuadam-performance},\eqref{offload-mem}$\sim$\eqref{offload-bwd} & 35.25 & 242.00 & 22.79 & 17.44 & 1.98 & 2.72 & 0.63 & 0.02 \\
 & 2048 & 8 & 1 & 8 & 128 & 2 & \eqref{suanzi-up}$\sim$\eqref{tp-overlap-costmodel} & \eqref{PPOverlap} & \eqref{dpoverlap} & \eqref{distributed-optimizer-memory},\eqref{distributed-optimizer-performance},\eqref{offload-mem}$\sim$\eqref{offload-bwd} & 34.46 & 234.16 & 23.55 & 17.40 & 0.50 & 2.72 & 0.69 & 2.25 \\
 & 2048 & 8 & 1 & 16 & 64 & 2 & \eqref{suanzi-up}$\sim$\eqref{tp-overlap-costmodel} & \eqref{PPOverlap} & \eqref{dpoverlap} & \eqref{cpuadam-mem},\eqref{cpuadam-performance},\eqref{offload-mem}$\sim$\eqref{offload-bwd} & 22.78 & 224.78 & 24.53 & 18.26 & 0.50 & 5.47 & 0.29 & 0.01 \\
 & 8192 & 8 & 1 & 8 & 128 & 2 & \eqref{suanzi-up}$\sim$\eqref{tp-overlap-costmodel} & \eqref{PPOverlap} & \eqref{dpoverlap} & \eqref{cpuadam-mem},\eqref{cpuadam-performance},\eqref{offload-mem}$\sim$\eqref{offload-bwd} & 35.25 & 320.39 & 68.85 & 63.59 & 1.83 & 2.72 & 0.69 & 0.02 \\
 & 8192 & 8 & 1 & 8 & 128 & 2 & \eqref{suanzi-up}$\sim$\eqref{tp-overlap-costmodel} & \eqref{PPOverlap} & \eqref{dpoverlap} & \eqref{distributed-optimizer-memory},\eqref{distributed-optimizer-performance},\eqref{offload-mem}$\sim$\eqref{offload-bwd} & 34.46 & 310.30 & 71.09 & 63.59 & 1.83 & 2.72 & 0.69 & 2.25 \\
 & 8192 & 4 & 1 & 16 & 128 & 2 & \eqref{suanzi-up}$\sim$\eqref{tp-overlap-costmodel} & \eqref{PPOverlap} & \eqref{dpoverlap} & \eqref{cpuadam-mem},\eqref{cpuadam-performance},\eqref{offload-mem}$\sim$\eqref{offload-bwd} & 36.74 & 304.53 & 72.43 & 65.18 & 0.83 & 5.77 & 0.63 & 0.02 \\
 & 8192 & 8 & 1 & 16 & 64 & 2 & \eqref{suanzi-up}$\sim$\eqref{tp-overlap-costmodel} & \eqref{PPOverlap} & \eqref{dpoverlap} & \eqref{cpuadam-mem},\eqref{cpuadam-performance},\eqref{offload-mem}$\sim$\eqref{offload-bwd} & 22.78 & 301.26 & 73.22 & 65.61 & 1.84 & 5.47 & 0.29 & 0.01 \\
\hline
\end{tabular}}
\end{table}

%% file: Table/5/optimizereffect.tex
\begin{table}[]
\centering
\caption{Effect of Optimizer Type on single DP group}
\renewcommand{\arraystretch}{1.3}
\label{tab:optimizer-effect}
\scalebox{0.9}{
\begin{tabular}{ccccccccc}
\hline
\textbf{$G_n$} & \textbf{$G_{bs}$} & $t$ & $c$ & $p$ & $d$ & Memory & TFLOPS & Optimizer Type \\ \hline
64 & 16 & 8 & 1 & 8 & 1 & 34.46 & 241.18 & CPU Optimizer \\
128 & 16 & 8 & 1 & 16 & 1 & 21.71 & 178.88 & CPU Optimizer \\
256 & 16 & 8 & 1 & 32 & 1 & 43.50 & 120.63 & Distributed Optimizer \\ \hline
\end{tabular}}
\end{table}

%% file: Table/5/N-effect.tex
\begin{table}[h!]
\centering
\caption{Impact of $N$ on optimal $e$.}
\label{tab:N-effect}
\renewcommand{\arraystretch}{1.3}
\scalebox{0.75}{
\begin{tabular}{cccccccccccccc}
\hline
$G_n$                    & $N$   & $G_{bs}$   & $t$ & $e$  & $d$  & $p$ & $m_{bs}$ & $T^\mathrm{step}$ & $T^\mathrm{cal}$ & $T^\mathrm{EP}$ & $T^\mathrm{PP}$ & $T^\mathrm{DP}$ & $T^\mathrm{update}$ \\ \hline
\multirow{6}{*}{1024} & 8   & 7680  & 1  & 8   & 16  & 8  & 2 & 58.95    & 46.80   & 2.30   & 0.57   & 3.50   & 5.79       \\
                      & 16  & 7680  & 1  & 16  & 8   & 8  & 2 & 53.55    & 45.12   & 2.55   & 0.57   & 1.32   & 3.98       \\
                      & 32  & 7680  & 1  & 32  & 4   & 8  & 2 & 51.93    & 44.28   & 2.84   & 0.57   & 1.17   & 3.08       \\
                      & 64  & 7680  & 1  & 64  & 4   & 4  & 2 & 51.25    & 41.18   & 3.15   & 0.57   & 1.86   & 4.49       \\
                      & 128 & 7680  & 1  & 128 & 1   & 8  & 2 & 50.50    & 43.76   & 3.50   & 0.57   & 0.27   & 2.40       \\
                      & 256 & 7680  & 1  & 128 & 2   & 4  & 2 & 49.39    & 41.04   & 3.50   & 0.57   & 0.47   & 3.81       \\ \hline
\multirow{6}{*}{2048} & 8   & 15360 & 1  & 8   & 32  & 8  & 2 & 59.43    & 46.80   & 2.30   & 0.57   & 3.97   & 5.79       \\
                      & 16  & 15360 & 1  & 16  & 16  & 8  & 2 & 53.72    & 45.12   & 2.55   & 0.57   & 1.49   & 3.98       \\
                      & 32  & 15360 & 1  & 32  & 8   & 8  & 2 & 52.08    & 44.28   & 2.84   & 0.57   & 1.32   & 3.08       \\
                      & 64  & 15360 & 1  & 64  & 8   & 4  & 2 & 51.48    & 41.18   & 3.15   & 0.57   & 2.09   & 4.49       \\
                      & 128 & 15360 & 1  & 128 & 4   & 4  & 2 & 50.89    & 41.04   & 3.50   & 0.57   & 1.97   & 3.81       \\
                      & 256 & 15360 & 1  & 128 & 2   & 8  & 2 & 50.55    & 43.76   & 3.50   & 0.57   & 0.31   & 2.40       \\ \hline
\multirow{6}{*}{4096} & 8   & 30720 & 1  & 8   & 64  & 8  & 2 & 59.92    & 46.80   & 2.30   & 0.57   & 4.46   & 5.79       \\
                      & 16  & 30720 & 1  & 16  & 32  & 8  & 2 & 54.94    & 45.12   & 2.55   & 0.57   & 2.71   & 3.98       \\
                      & 32  & 30720 & 1  & 32  & 16  & 8  & 2 & 52.24    & 44.28   & 2.84   & 0.57   & 1.47   & 3.08       \\
                      & 64  & 30720 & 1  & 64  & 8   & 8  & 2 & 51.64    & 43.90   & 3.15   & 0.57   & 1.38   & 2.63       \\
                      & 128 & 30720 & 1  & 128 & 8   & 4  & 2 & 51.13    & 41.04   & 3.50   & 0.57   & 2.21   & 3.81       \\
                      & 256 & 30720 & 1  & 256 & 4   & 4  & 2 & 51.12    & 41.04   & 3.89   & 0.57   & 2.15   & 3.47       \\ \hline
\multirow{6}{*}{8192} & 8   & 61440 & 1  & 8   & 128 & 8  & 2 & 60.45    & 46.80   & 2.30   & 0.57   & 4.99   & 5.79       \\
                      & 16  & 61440 & 1  & 16  & 64  & 8  & 2 & 55.27    & 45.12   & 2.55   & 0.57   & 3.04   & 3.98       \\
                      & 32  & 61440 & 1  & 32  & 32  & 8  & 2 & 52.41    & 44.28   & 2.84   & 0.57   & 1.64   & 3.08       \\
                      & 64  & 61440 & 1  & 64  & 16  & 8  & 2 & 51.80    & 43.90   & 3.15   & 0.57   & 1.54   & 2.63       \\
                      & 128 & 61440 & 1  & 128 & 16  & 4  & 2 & 51.38    & 41.04   & 3.50   & 0.57   & 2.46   & 3.81       \\
                      & 256 & 61440 & 1  & 256 & 8   & 4  & 2 & 51.36    & 41.04   & 3.89   & 0.57   & 2.39   & 3.47       \\ \hline
\end{tabular}}
\end{table}

%% file: Table/5/roofline-profile.tex
\begin{table}[t]
\centering
\caption{profile-based tuning vs roofline-based tuning}
\label{tab:profile-roofline}
\renewcommand{\arraystretch}{1.3}
\scalebox{0.8}{
\begin{tabular}{ccccccccccccccc}
\hline
$G_n$ & $G_{bs}$ & $t$ & c & p & $d$ & $m_{bs}$ & $v$ & TFLOPS & $T^\mathrm{step}$ & $T^\mathrm{cal}$ & $T^\mathrm{TP}$ & $T^\mathrm{PP}$ & $T^\mathrm{DP}$ & $T^\mathrm{update}$ \\ \hline
\multirow{4}{*}{64} & \multirow{2}{*}{16} & 8 & 1 & 8 & 1 & 2 & 8 & 241.18 & 22.86 & 17.40 & 0.50 & 2.72 & 0.00 & 2.25 \\
 &  & 8 & 1 & 8 & 1 & 2 & 8 & 247.10 & 22.32 & 16.85 & 0.50 & 2.72 & 0.00 & 2.25 \\ \cline{2-15} 
 & \multirow{2}{*}{64} & 8 & 1 & 8 & 1 & 2 & 8 & 313.33 & 70.40 & 63.59 & 1.83 & 2.72 & 0.00 & 2.25 \\
 &  & 8 & 1 & 8 & 1 & 2 & 8 & 322.62 & 68.37 & 61.57 & 1.83 & 2.72 & 0.00 & 2.25 \\ \hline
\multirow{4}{*}{2048} & \multirow{2}{*}{512} & 8 & 1 & 8 & 32 & 2 & 8 & 260.60 & 21.16 & 17.40 & 0.50 & 2.72 & 0.53 & 0.02 \\
 &  & 8 & 1 & 8 & 32 & 2 & 8 & 267.52 & 20.61 & 16.85 & 0.50 & 2.72 & 0.53 & 0.02 \\ \cline{2-15} 
 & \multirow{2}{*}{2048} & 8 & 1 & 8 & 32 & 2 & 8 & 321.10 & 68.70 & 63.59 & 1.83 & 2.72 & 0.53 & 0.02 \\
 &  & 8 & 1 & 8 & 32 & 2 & 8 & 330.87 & 66.67 & 61.57 & 1.83 & 2.72 & 0.53 & 0.02 \\ \hline
\multirow{4}{*}{8192} & \multirow{2}{*}{2048} & 8 & 1 & 8 & 128 & 2 & 8 & 258.73 & 21.31 & 17.40 & 0.50 & 2.72 & 0.69 & 0.02 \\
 &  & 8 & 1 & 8 & 128 & 2 & 8 & 264.28 & 20.87 & 16.85 & 0.50 & 2.72 & 0.79 & 0.02 \\ \cline{2-15} 
 & \multirow{2}{*}{8192} & 8 & 1 & 8 & 128 & 2 & 8 & 320.39 & 68.85 & 63.59 & 1.83 & 2.72 & 0.69 & 0.02 \\
 &  & 8 & 1 & 8 & 128 & 2 & 8 & 329.62 & 66.92 & 61.57 & 1.83 & 2.72 & 0.79 & 0.02 \\ \hline
\end{tabular}}
\end{table}

%% file: futurework.tex
\section{Future Work}
This paper focuses on the design of MoFa and systematic analysis of tuning results, but there are still many challenges that remain unsolved. We summarize the work that we will continue to explore in the future as follows

\textbf{Modeling Guidance and Digital Twin Simulation}: Providing modeling strategies to guide the development of minimal systems and employing simulation-based digital twins for validation within training clusters.

\textbf{E2E Optimal Strategy Optimization}: Investigating the differences in hardware failure probabilities under various step-based optimization strategies and identifying the optimal ETTR (Expected Time to Repair) for cluster efficiency, thereby determining the best end-to-end training strategy.

\textbf{Optimal Hardware Configuration}: For a given training task (e.g., model size, dataset scale, and expected training duration), modeling and recommending the optimal training strategy along with the corresponding hardware configuration (GPU type, number of cards, and network architecture).

\textbf{Optimal Model Configuration}: For a given hardware cluster (GPU type, number of cards, and network architecture), modeling and determining the most suitable model structure and parameter scale.

%% file: conclusion.tex
\section{Conclusion}
This paper has presented \textbf{MoFa}, an innovative performance modeling framework that systematically addresses two critical limitations in large-scale LLM pretraining: the inadequate consideration of multi-dimensional optimization features and the neglect of fault tolerance overhead. Our framework's core innovation lies in its unified approach that seamlessly integrates an enhanced cost model for capturing key optimization effects with a historical reliability-aware fault tolerance model. This integrated design enables MoFa to achieve remarkable prediction accuracy across diverse training scenarios. Through extensive tuning experiments, our framework has systematically revealed the intricate relationships between parallelization strategies, optimization techniques, and system reliability, providing unprecedented insights into the key factors governing pretraining performance at scale.

%% file: author.tex
\section{Authors}
The list of authors is shown below. Names marked with an asterisk $^*$ indicate the equal contributor, names marked with an asterisk $^\dagger$ indicate the corresponding authors.

Lu Zhao$^*$, Rong Shi$^*$, Shaoqing Zhang$^{*\dagger}$, Shangchao Su, Ziqing Yin, Zhiyan Cui, Hongfeng Sun, Baoguo He, Yueqiang Chen, Liang Dong, Xiyuan Li, Lingbin Wang, Lijun Ma, Qiang Huang, Ting Liu, Chong Wang, Can Wei.